\documentclass[journal]{IEEEtran}
\usepackage[T1]{fontenc}
\usepackage{multirow}
\usepackage{psfrag}
\usepackage{longtable}
\usepackage{supertabular} 
\usepackage{graphicx}
\usepackage{lineno,hyperref}
\usepackage{amsmath}
\usepackage{amsfonts}
\usepackage{amssymb}
 \usepackage{flushend}
\usepackage{epstopdf}
\usepackage{array,multirow,makecell,booktabs}
\modulolinenumbers[5]
\usepackage{lscape}
\usepackage{rotating, adjustbox}
\usepackage[cmintegrals]{newtxmath}
\usepackage[table]{xcolor}
\usepackage{tikz}
\usetikzlibrary{arrows,shapes,positioning,shadows,trees}
\usetikzlibrary{trees}
	
\ifCLASSINFOpdf
\else
\fi
\newcolumntype{L}[1]{>{\raggedright\let\newline\\\arraybackslash\hspace{0pt}}m{#1}}
\newcolumntype{C}[1]{>{\centering\let\newline\\\arraybackslash\hspace{0pt}}m{#1}}
\newcolumntype{R}[1]{>{\raggedleft\let\newline\\\arraybackslash\hspace{0pt}}m{#1}}

\begin{document}
%
\title{Blockchain Technologies for the Internet of Things: Research Issues and Challenges}
%
%
%

\author{Mohamed~Amine~Ferrag, Makhlouf~Derdour, Mithun~Mukherjee,~\IEEEmembership{Member,~IEEE,} Abdelouahid Derhab, Leandros~Maglaras,~\IEEEmembership{Senior Member,~IEEE,} Helge Janicke 

\thanks{(Corresponding author: Mohamed Amine Ferrag)}
\thanks{M. A. Ferrag is with Department of Computer Science, Guelma University, 24000, Algeria, and also with Networks and Systems Laboratory (LRS), Badji Mokhtar-Annaba University, 23000 Annaba, Algeria e-mail: mohamed.amine.ferrag@gmail.com, ferrag.mohamedamine@univ-guelma.dz, phone: +213661-873-051}
\thanks{M. Derdour is with Department of Mathematics and Computer Science, University of Larbi Tebessi, Tebessa 12002, Algeria  e-mail: m.derdour@yahoo.fr}
\thanks{M. Mukherjee is with the Guangdong Provincial Key Laboratory of Petrochemical Equipment Fault Diagnosis, Guangdong University of Petrochemical Technology, Maoming 525000, China e-mail: m.mukherjee@ieee.org}
\thanks{A. Derhab is with Center of Excellence in Information Assurance (CoEIA), King Saud University, Saudi Arabia, e-mail: abderhab@ksu.edu.sa}
\thanks{L. Maglaras is with School of Computer Science and Informatics, De Montfort University, Leicester, UK, , and also with General Secretariat of Digital Policy, Athens, Greece, e-mail:  leandrosmag@gmail.com}
\thanks{H. Janicke is with School of Computer Science and Informatics, De Montfort University, Leicester, UK, e-mail: heljanic@dmu.ac.uk}
\thanks{Manuscript received 2018.}}

\maketitle

\begin{abstract}
This paper presents a comprehensive survey of the existing blockchain protocols for the Internet of Things (IoT) networks. We start by describing the blockchains  and summarizing the existing surveys that deal with blockchain technologies. Then, we provide an overview of the application domains of blockchain technologies in IoT, e.g, Internet of Vehicles, Internet of Energy, Internet of Cloud, Fog computing, etc. Moreover, we provide a classification of threat models, which are considered by blockchain protocols in IoT networks, into five main categories, namely, identity-based attacks, manipulation-based attacks, cryptanalytic attacks, reputation-based attacks, and service-based attacks. In addition, we provide a taxonomy and a side-by-side comparison of the  state-of-the-art methods towards secure and privacy-preserving blockchain technologies with respect to the blockchain model, specific security goals, performance, limitations, computation complexity, and communication overhead. Based on the current survey, we highlight open research challenges and discuss possible future research directions in the blockchain technologies for IoT. \end{abstract}


%
\IEEEpeerreviewmaketitle

\section{Introduction}

\begin{figure}[h]
\centering
\includegraphics[width=1\linewidth]{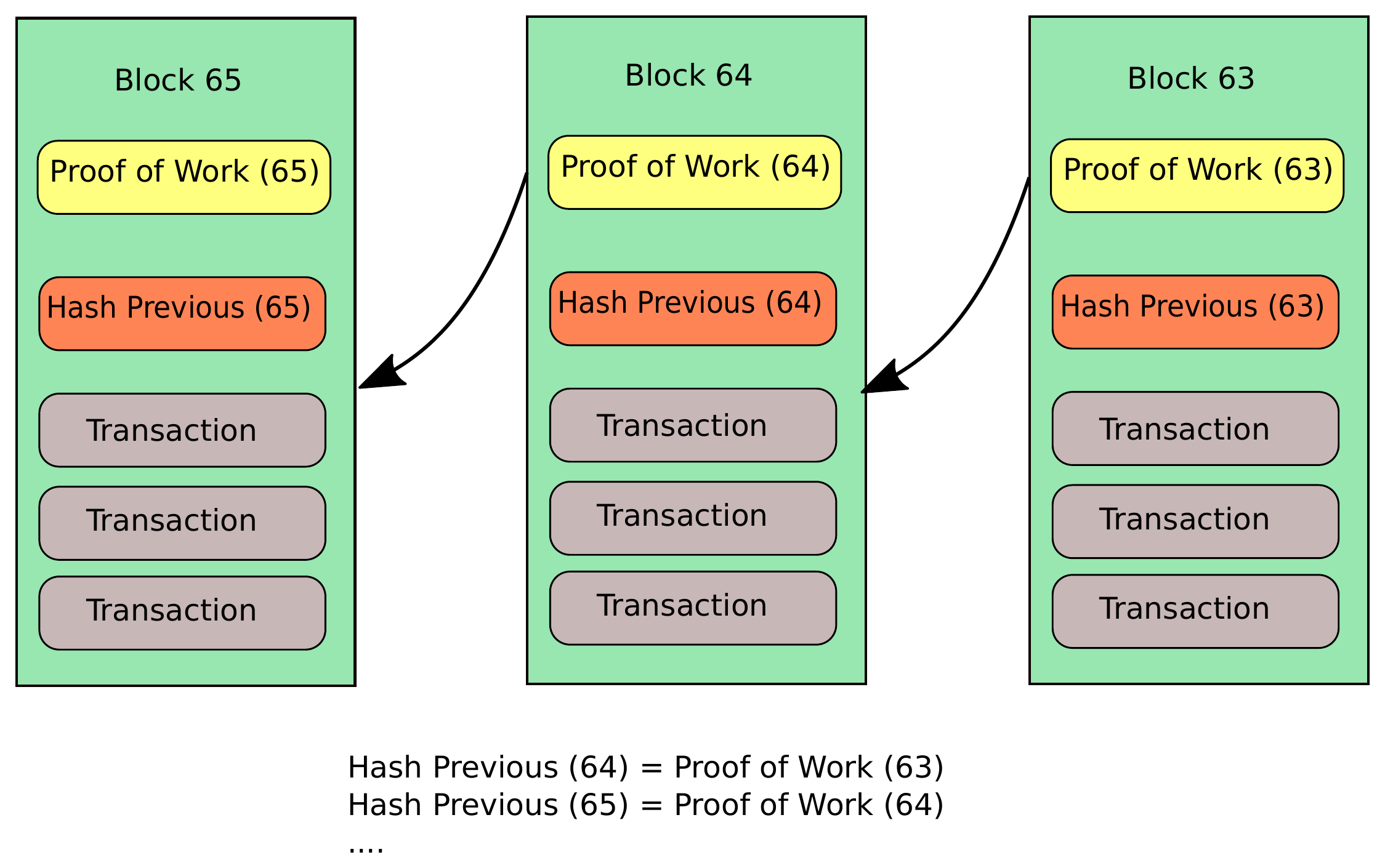}
\caption{Blockchain structure.}
\label{fig:Fig2}
\end{figure}

In the last few years, we have witnessed the potential of Internet of Things to deliver exciting services across several sectors, from social media, business, intelligent transportation and smart cities to the industries~\cite{IDCIoT2, IDCIoT, Miorandi2012internet}. IoT seamlessly interconnects heterogeneous devices with diverse functionalities in the human-centric and machine-centric networks to meet the evolving requirements of the earlier mentioned sectors. Nevertheless, the significant number of connected devices and massive data traffic become the bottleneck in meeting the required Quality-of-Services (QoS) due to the computational, storage, and bandwidth-constrained IoT devices. Most recently, the blockchain~\cite{Puthal2018CM, 51, SwanBlockchainBook, Tschorsch2016}, a paradigm shift, is transforming all the major application areas of IoT by enabling a decentralized environment with anonymous and trustful transactions. Combined with the blockchain technology, IoT systems benefit from the lower operational cost, decentralized resource management, robustness against threats and attacks, and so on. Therefore, the convergence of IoT and blockchain technology aims to overcome the significant challenges of realizing the IoT platform in the near future.  

Blockchain, a distributed append-only public ledger technology, was initially intended for the cryptocurrencies, e.g., Bitcoin\footnote{Apart from Bitcoin, there are several cryptocurrencies such as Litecoin, Peercoin, Swiftcoin, Peercoin, and Ripple.}.   
In 2008, Satoshi Nakamato~\cite{52} introduced the concept of blockchain that has attracted much attention over the past years as an emerging peer-to-peer (P2P)  technology for distributed computing and decentralized data sharing. Due to the adoption of cryptography technology and without a centralized control actor or a centralized data storage, the blockchain can avoid the attacks that want to take control over the system. Later, in 2013, Ethereum,  a transaction-based state-machine, was presented to program the blockchain technologies. Interestingly, due to its unique and attractive features such as: transactional privacy, security, the immutability of data, auditability, integrity, authorization, system transparency, and fault tolerance, blockchain is being applied in several sectors beyond the cryptocurrencies. Some of the areas are identity management~\cite{Wilson2015}, intelligent transportation~\cite{3,13,20,23,31,36}, supply-chain management, mobile-crowd sensing~\cite{27}, agriculture~\cite{FengTian2016}, Industry 4.0~\cite{32, Ahram2017}, Internet of energy~\cite{4,25,32,33}, and security in mission critical systems~\cite{Kshetri2017}.

\begin{figure}
	\centering
	\includegraphics[width=.9\columnwidth]{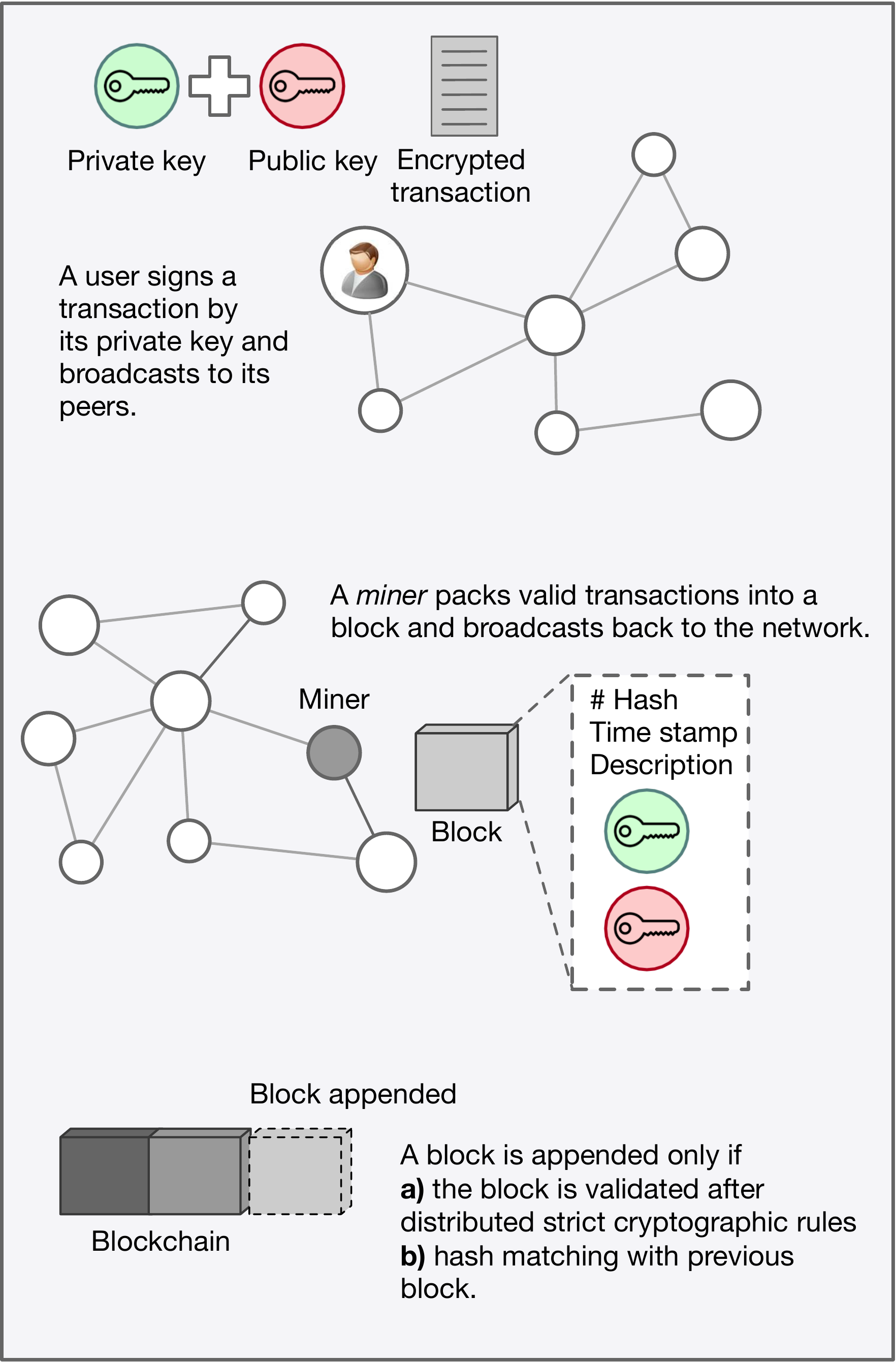}
	\caption{An illustration of blockchain working methodology.}
	\label{fig:BlockChain}%
\end{figure}

As shown in Fig. \ref{fig:Fig2}, the blockchain structure is composed of a sequence of blocks,  which are linked together by their hash values. In the blockchain network, a public ledger maintains the digitally signed transactions of the users in a P2P network. In general, a user has two keys: a public key for other users for the encryption and a private key to read an encrypted message, as shown in Fig. \ref{fig:BlockChain}. From the blockchain perspective, the private key is used for signing the blockchain transaction and the public key represents the unique address. Asymmetric cryptography is used to decrypt the message encrypted by the corresponding  public key. At the initial stage, a user signs a transaction using its private key and broadcasts it to its peers. Once the peers receive the signed transaction, they validate the transaction and disseminate it over the network. All the parties who are involved in the transactions mutually validate the transaction to meet a consensus agreement. Once a distributed consensus is reached, the special node, called as miners, includes the valid transaction into a timestamped block. The block, which is included by the miner, is broadcast back into the network. After validating  the broadcast block, which  contains the transaction, as well as hash-matching it with the previous block in the blockchain, the broadcast block is appended to the blockchain. 
%
%
%

\begin{figure}[t!]
\centering
\includegraphics[width=0.8\linewidth]{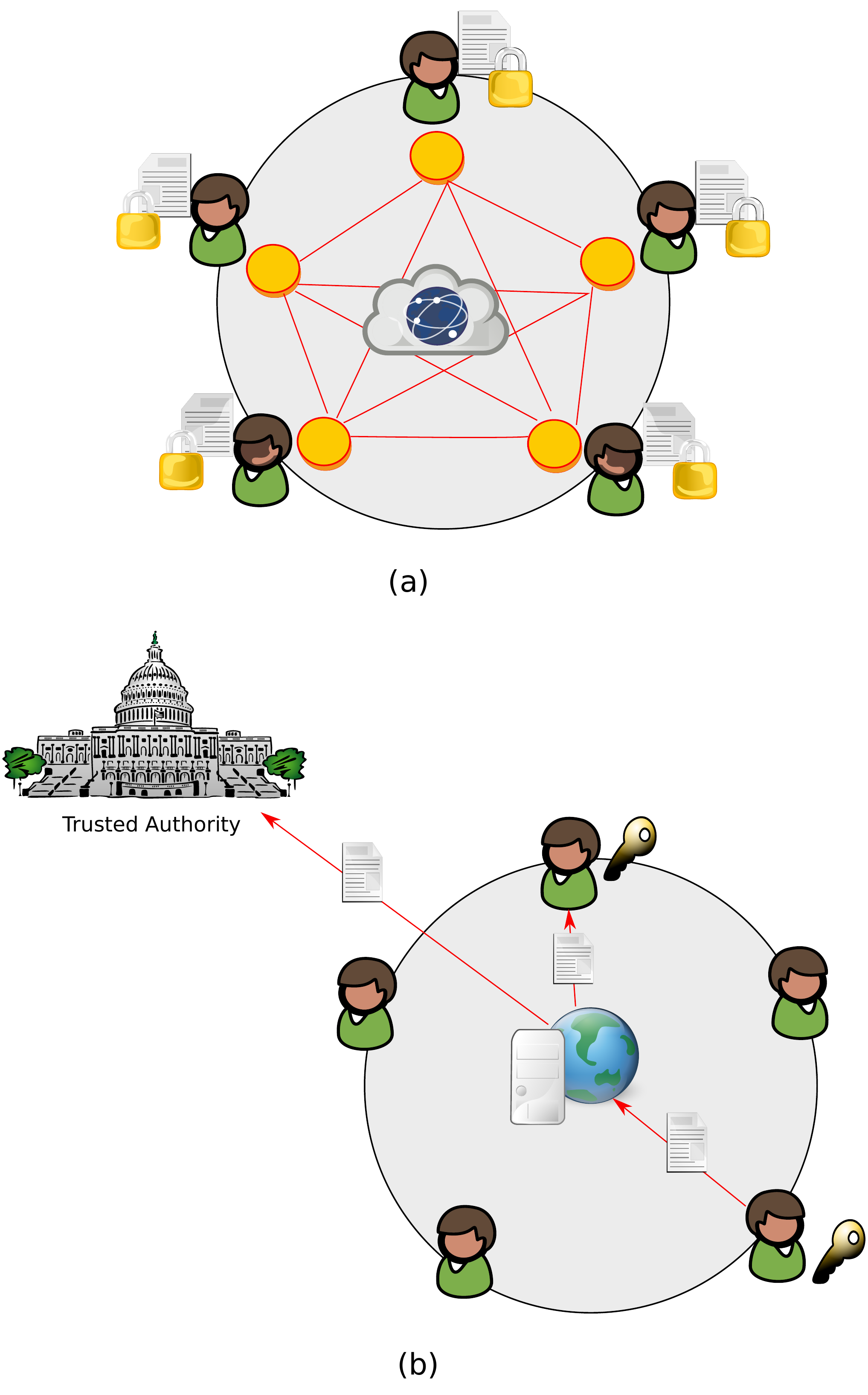}
\caption{(a) Public blockchain system; (b) Private blockchain system.}
\label{fig:Fig3}
\end{figure}

Based on the data management and the type of  applications, blockchain can classified either as private (permission) or public (permissionless). Both classes are decentralized and provide a certain level of immunity against faulty or malicious users for the ledger. The main differences between private and public blockchains lie in the execution of the consensus protocol, the maintenance of the ledger, and the authorization to join to the P2P network. Detailed examples of these classes are illustrated in~\cite{FernandezCarames2018}. In the context of IoT, blockchains can be classified based on authorization and authentication. As shown in Fig.~\ref{fig:Fig3}, in a private blockchain, the centralized trusted authority that manages the authentication and authorization process selects the miners. On the other hand, in a public blockchain (in general, permissionless), there is no intervention of any third-party for the miner selection and joining for a new user to the blockchain network.  

Recently, there is a huge amount of investment from the industries~\cite{CryptocurrencyMarket, BlockchainReport} as well as a significant interest from academia to solve major research challenges in blockchain technologies. For example, the consensus protocols are the major building blocks of the blockchain  technologies, thus, the  threats targeting  the consensus protocols become a significant research issue in the blockchain. Furthermore, blockchain forks bring threats to the blockchain consensus protocols. Moreover, it is observed that the vulnerability is about 51\% for a new blockchain~\cite{Bitcoin51attack}. At the same time,  maintenance of several blockchains requires a significant amount of power consumption~\cite{Muneeb2016}. 

\subsection{Related Surveys and Our Contributions} 

There are related survey papers~\cite{Tschorsch2016, Sankar2017, Kaushik2017, Khalilov2018, FernandezCarames2018, Conti2018} that covered different aspects of the blockchain technology. For example, a brief overview of blockchain for bitcoin was discussed in~\cite{Sankar2017, Kaushik2017}. However, these surveys are very limited regarding detailed discussion on research challenges in blockchain.
Moreover, Sankar et al.~\cite{Sankar2017} briefly presented the feasibility   of various consensus protocols in the blockchain. The detailed insights of bitcon were presented in~\cite{Tschorsch2016}. Recently, the surveys~\cite{FernandezCarames2018} presented the overview of Blockchain-based IoT (BIoT) applications. The security and privacy aspects are presented in~\cite{Conti2018, Khalilov2018} for bitcoin, one of the blockchain applications. Table~\ref{Table:RelatedSurvey} summarizes the main focuses and major contributions of the previous comprehensive surveys on blockchain technologies.
Although the above mentioned  surveys~\cite{Tschorsch2016, Khalilov2018, FernandezCarames2018, Conti2018} have laid a solid foundation for blockchain technologies, our survey differs in several aspects. The main contributions of this paper are:
\begin{itemize}
\item We provide overviews of the different application domains of blockchain technologies in IoT, e.g, Internet of Vehicles, Internet of Energy, Internet of Cloud, Fog computing, etc. 
\item We classify the threat models, which are considered by the blockchain protocols in IoT networks, into five main categories, namely, identity-based attacks, manipulation-based attacks, cryptanalytic attacks, reputation-based attacks, and service-based attacks.
\item We review existing research on anonymity and privacy in Bitcoin systems.
\item We provide a taxonomy and a side-by-side comparison, in a tabular form, of the state-of-the-art on the recent advancements towards secure and privacy-preserving blockchain technologies with respect to blockchain model, specific security goals, performance and limitations, computation complexity and communication overhead.
\item We highlight the open research challenges and discuss the possible future research directions  in the field of  blockchain technologies for IoT.
\end{itemize}

\begin{table}[!t]
	\centering
	\caption{Related Surveys on Blockchain Technologies}
	\label{Table:RelatedSurvey}
	\setlength{\tabcolsep}{4pt}
	\vspace*{-\baselineskip}
	\renewcommand{\arraystretch}{1.2}
	\begin{tabular}{L{0.3in} L{0.8in} L{1.8in}}\toprule
		\rowcolor{black!15}   Year & Author & Main focus/contributions \\\midrule
		2016	& Tschorsch and Scheuermann~\cite{Tschorsch2016}	&  Fundamental structures and insights of the core of the Bitcoin protocol and its applications\\\hline
		\rowcolor{black!5}2017 & Sankar et al.~\cite{Sankar2017}&  Feasibility and efficiency of consensus protocols in blockchain.\\\hline
		2017 & Kaushik et al.~\cite{Kaushik2017} & A brief survey on bitcoin.\\\hline
		\rowcolor{black!5}2018 & Khalilov and  Levi~\cite{Khalilov2018} &  An overview and detailed investigation of anonymity and privacy in Bitcoin-like digital cash systems.\\\hline
		2018 & Fern\'{a}ndez-Caram\'{e}s and Fraga-Lamas~\cite{FernandezCarames2018}&  A review on developing Blockchain-based IoT (BIoT) applications.\\\hline		
		\rowcolor{black!5}2018 & Conti et al.~\cite{Conti2018} & A systematic survey that covers the security and the privacy aspects of Bitcoin.\\	
		\bottomrule     
	\end{tabular}
\end{table}

The remainder of this paper is organized as follows. Section \ref{sec:2}  presents the application domains of blockchain technologies in IoT. In Section  \ref{sec:3}, we present the classification of threat models that are considered by the blockchain protocols in IoT networks. In Section \ref{sec:4}, we present a side-by-side comparison, in a tabular form, of the  state-of-the-art on the recent advancements towards secure and privacy-preserving blockchain technologies. Then, we discuss open issues and recommendations for further research in Section \ref{sec:5}. Finally, we draw our conclusions in Section \ref{sec:6}.

\begin{figure*}[t!]
\centering
\includegraphics[width=1\linewidth]{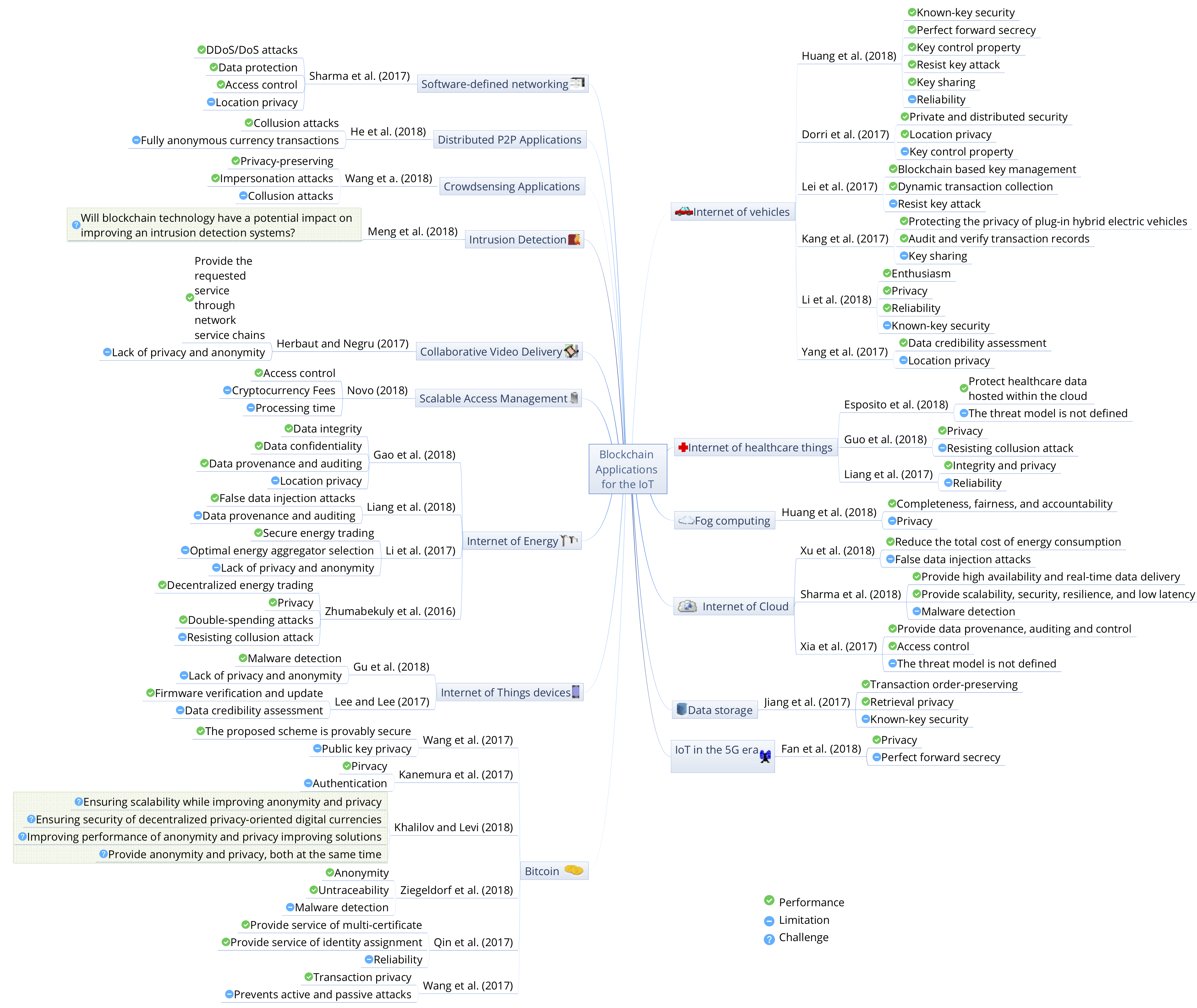}
\caption{Blockchain applications for the IoT.}
\label{fig:Fig1}
\end{figure*}
\section{Blockchain applications for the IoT} \label{sec:2}
As presented in Fig. \ref{fig:Fig1}, the blockchain technology can be effectively applied in almost all domains of IoT.
\subsection{Internet of healthcare things}
The usage of IoT in healthcare has allowed to feed the ehealthcare systems with  clinical data related to the  patients, their family, their friends, as well as the healthcare providers. The data, called electronic medical records (EMRs),  is stored by the responsible healthcare provider. To facilitate patient data portability, there are the electronic health records (EHRs), which have a richer data structure than EMRs. Based on the idea of distributed online database, Esposito et al. \cite{1} proposed the design of a blockchain-based scheme for the IoT in healthcare. In a model of consortium blockchain, a new block is instantiated and distributed when new healthcare data is created. To preserve the privacy of patients and maintain the immutability of EHRs, Guo et al. \cite{7} introduced an attribute-based signature scheme, named MA-ABS, which uses multiple authorities. The MA-ABS scheme uses the blockchain technology and can resist to N-1 corrupted authorities collusion attacks. In addition, the MA-ABS is unforgeable in suffering a selective predicate attack. Therefore, Liang et al. \cite{35} used the blockchain network in mobile healthcare applications for integrity protection, further auditing or investigation.
\begin{table*}[!t]
	\centering
	\caption{Major attacks on Blockchain}
	\label{Table:Tab3}
	\setlength{\tabcolsep}{2pt}
		\vspace*{-\baselineskip}
		\renewcommand{\arraystretch}{1.5}
{\scriptsize
\begin{tabular}{p{1.2in}|p{3.2in}|p{1.5in}} \toprule
\rowcolor{black!15}
\textbf{Threat model} & \textbf{Countermeasures} & \textbf{Resistant protocols} \\ \midrule
Key attack & - Elliptic curve encryption is used to calculate the hash functions & LNSC protocol \cite{3} \\ \hline 
\multirow{4}{*}{DDoS/DoS attack} & - Distributed SDN architecture & DistBlockNet protocol \cite{22} \\ 
 & - Decentralized mixing service & CoinParty protocol \cite{43} \\  
 & - Ring signature using ECDSA & Liu et al.'s protocol \cite{55} \\ 
 & -Block size limitation,  attribute-based signatures, and multi-receivers encryption & BSeIn protocol \cite{57} \\ \hline 
\multirow{2}{*} {Replay attack} & - Elliptic curve encryption is used to calculate the hash functions & LNSC protocol \cite{3} \\
 & - The freshness of public/private key pairs & BSeIn protocol \cite{57} \\ \hline 
Hiding Blocks & - An immutable chain of temporally ordered interactions is created for each agent & TrustChain protocol \cite{40} \\ \hline 
False data injection attack & - Blockchain consensus mechanisms & Liang et al.'s protocol \cite{25} \\ \hline 
Tampering attack & - Public-key cryptosystem & Wang et al.'s protocol \cite{45} \\ \hline 
\multirow{3}{*} {Impersonation attack} & - Elliptic curve encryption is used to calculate the hash functions & LNSC protocol \cite{3} \\  
 & - Distributed incentive mechanism based blockchain and the node cooperation based privacy protection mechanism & Wang et al.'s protocol \cite{27} \\ 
 & - Attribute-based signatures & BSeIn protocol \cite{57} \\ \hline 
Refusal to Sign & - Not interacting with the malicious agent, or splitting the transactions in smaller amounts & TrustChain protocol \cite{40} \\ \hline 
Overlay attack & - Every transaction is embedded with a Time-Stamp to mark the uniqueness & Wang et al.'s protocol \cite{45} \\ \hline 
\multirow{2}{*} {Double-spending attack} & - Multi signatures and anonymous encrypted message propagation streams & Aitzhan and Svetinovic's protocol \cite{33} \\ 
 & - Time-Stamp and the Proof-of-Work mechanism & Wang et al.'s protocol \cite{45} \\ \hline 
\multirow{2}{*} {Modification attack} & - Elliptic curve encryption is used to calculate the hash functions & LNSC protocol \cite{3} \\ 
 & - The attribute signature and the MAC & BSeIn protocol \cite{57} \\ \hline 
Collusion attack & - Blockchain-based incentive mechanism & He et al.'s protocol \cite{24} \\ \hline 
Whitewashing attack & - Lower priorities are given to the agents of new identities & TrustChain protocol \cite{40} \\ \hline 
Quantum attack & - Lattice-based signature scheme & Yin et al.'s protocol \cite{11} \\ \hline 
\multirow{2}{*} {Man-in-the-middle attack} & - Elliptic curve encryption is used to calculate the hash functions & LNSC protocol \cite{3} \\ 
 & - Secure mutual authentication & BSeIn protocol \cite{57} \\ \hline 
Sybil attack & - An immutable chain of temporally ordered interactions is created for each agent & TrustChain protocol \cite{40} \\ 
\bottomrule
\end{tabular}
}
\end{table*}

\subsection{Internet of things in the 5G era}

In the IoT era, 5G will enable a fully mobile and connected society for  billions of connected objects \cite{48}. To solve the privacy issues in the 5G heterogeneous communication environment, Fan et al. \cite{2} proposed a blockchain-based privacy preserving and data sharing scheme. Based on the idea of adding blocks to the blockchain, each new block is connected to the blockchain by its hash value. Note that the previous hash value can be known from the block header. 

\subsection{Internet of vehicles}

The Internet of Vehicles (IoV) is an emerging concept, which allows the integration of vehicles into the new era of the IoT in order to establish the smart communication between vehicles and heterogeneous networks such as vehicle-to-vehicle, vehicle-to-road, vehicle-to-human, vehicle-to-sensor, and vehicle-to-everything. However, some recent works try to apply the blockchain technology to  IoV. Based on the decentralized security model, Huang et al. \cite{3} proposed a blockchain ecosystem model, named LNSC, for electric vehicle and charging pile management. The LNSC model uses elliptic curve cryptography (ECC) to calculate hash functions electric vehicles and charging piles. To avoid the location tracking in the IoV, Dorri et al. \cite{13} proposed a decentralized privacy-preserving architecture, where overlay nodes manage the blockchain. In addition, the hash of the backup storage is stored in the blockchain. 

Without the administration from the central manager, Lei et al. \cite{20} proposed a blockchain-based dynamic key management for vehicular communication systems. Based on a decentralized blockchain structure, the third-party authorities are removed and the key transfer processes are verified and authenticated by the security manager network. Moreover, Kang et al. \cite{23} introduced a P2P electricity trading system, named PETCON, to illustrate detailed operations of localized P2P electricity trading. Using consortium blockchain method, the PETCON system can publicly audit and share transaction records without relying on a trusted third party. To solve the issues of forwarding reliable announcements without revealing users' identities, Li et al. \cite{31} proposed a privacy-preserving scheme, named CreditCoin, for sending announcements anonymously in the IoV. The CreditCoin scheme uses the blockchain via an anonymous vehicular announcement aggregation protocol to build trust in the IoV communications. For data credibility assessment in the IoV, Yang et al. \cite{36} proposed a blockchain-based reputation system, which can judge the received messages as either true or false based on the senders' reputation values.

\subsection{Internet of Energy}
The Internet of energy (IoE) provides an innovative concept to increase the visibility of energy consumption in the Smart Grid. Based on the sovereign blockchain technology, Gao et al. \cite{4} introduced a monitoring system on Smart Grid, named GridMonitoring, for ensuring transparency, provenance, and immutability. The GridMonitoring system is based on four layers, namely, 1) Registration and authentication layer, 2) Smart meter, 3) Processing and consensus nodes, and 4) Data processing on the smart grid network. In modern power systems, Liang et al. \cite{25} proposed a data protection framework based on distributed blockchain, which can resist against data manipulation that are launched by cyber attackers (e.g., false data injection attacks). To guarantee data accuracy, Liang's framework uses the consensus mechanism, which is automatically implemented by every node and has the representative characteristics, namely, 1) Setting of public/private key update frequency, 2) Block generation, 3) Miner selection, and 4) Release of meter's memory periodically. For secure energy trading in Industrial Internet of Things (IIoT), Li et al. \cite{32} introduced the energy blockchain, which is based on the consortium blockchain technology and the Stackelberg game. Aitzhan and Svetinovic \cite{33} implemented a token-based private decentralized energy trading system for in decentralized smart grid energy, which can be applied to the IoE.
\subsection{Internet of Things devices}
In the Internet of Things devices, attackers seek to exfiltrate the data of IoT devices by using the malicious codes in malware, especially on the open source Android platform. By utilizing statistical analysis method, Gu et al. \cite{5} introduced a malware detection system based on the consortium Blockchain, named CB-MDEE, which is composed of detecting consortium chain by test members and public chain by users. The CB-MDEE system adopts a fuzzy comparison method and multiple marking functions In order to reduce the false-positive rate and improve the detection ability of malware variants. To protect the embedded devices in the IoT, Lee et al. \cite{47} a firmware update scheme based on the blockchain technology, which the embedded devices have the two different operation cases, namely, 1) response from a verification node to a request node, and 2) response from a response node to a request node.

\subsection{Access Management in IoT}
For managing IoT devices, Novo \cite{8} proposed a distributed access control system using the blockchain technology. The architecture of this system is composed of six components, namely, 1) Wireless sensor networks, 2) Managers, 3) Agent node, 4) Smart contract, 5) Blockchain network, and 6) Management hubs. This system brings some advantages for the access control in IoT, such as: 1) mobility, which can be used in isolated administrative systems; 2) accessibility, which ensures that the access control rules are available at any time; 3) concurrency, which allows that the access control policies can be modified simultaneously; 4) lightweight, which means that the IoT devices do not need any modification to adopt this system; 5) scalability, as the IoT devices can be connected through different constrained networks; 6) transparency, where the system can preserve the location privacy.

\subsection{Collaborative video delivery}
The diffusion of high-quality content in the IoT nowadays challenges for the internet service providers. However, Herbaut and Negru \cite{17} proposed a decentralized brokering mechanism for collaborative blockchain-based video delivery, which is relying on advanced network services chains. Specifically, this management mechanism is composed of three blockchains, namely, 1) the content brokering blockchain, 2) the delivery monitoring blockchain, and 3) the provisioning blockchain. In addition, this management mechanism is deployed with the open source project Hyperledger-Fabric \footnote{www.hyperledger.org/projects/fabric} where the results show that the number of nodes slightly increases the convergence time .

\subsection{Internet of Cloud}
In the Internet of Cloud (IoC), billions of IoT devices upload their data to the cloud through the internet connection utilizing virtualization technology. Xu et al. \cite{10} introduced an intelligent resource management for cloud datacenters based on the blockchain technology, in order to save and reduce the total cost of energy consumption. Specifically, the users use their individual private keys to sign a transaction, while the neighboring users verify the broadcast transaction. The block is discarded when it does not pass verification. Therefore, Sharma et al. \cite{14} proposed a distributed cloud architecture that uses three emerging technologies, namely, software-defined networking (SDN), fog computing, and a blockchain technique. The SDN controllers of the fog node are used to provide programming interfaces to network management operators. The blockchain technique is used to provide scalable, reliable, and high-availability services. In addition, Xia et al. \cite{21} proposed a blockchain-based data sharing system, named MeDShare, for cloud service providers. This system uses four layers namely, 1) User layer, 2) Data query layer, 3) Data structuring and provenance layer, and 4) Existing database infrastructure layer.

\subsection{Intrusion Detection}
Many techniques for implementing intrusion detection systems (IDSs) in the IoT environment have been proposed, which are based in machine learning. To improve the collaborative intrusion detection systems (CIDSs), Alexopoulos et al. \cite{49} introduced the idea of utilizing blockchain technology in order to secure the exchange of alerts between the collaborating nodes. Meng et al. \cite{9} discussed the applicability of blockchain technology in a intrusion detection systems. Modern intrusion detection systems must be based on collaborative communication among distributed IDSs \cite{cruz2016cybersecurity}, demanding extensive data sharing among entities and trust computation. In order to deal with the privacy concerns that are raised by the data exchange and to suppress insider attacks, the blockchain technology is applied. In this way, the use of trusted third party, which is also a single point of failure, that is needed in traditional collaborative IDSs can be avoided.

\subsection{Software-defined networking}
To increase IoT's bandwidth, researchers have been proposing the Software Defined Networking (SDN) technology, which provides intelligent routing and simplifies decision-making processes by the SDN controller \cite{50}. Recently, Sharma et al. \cite{22} proposed a distributed IoT network architecture, named DistBlockNet. Based on the blockchain technology,  DistBlockNet architecture can provide scalability and flexibility, without the need for a central controller. The distributed blockchain network uses two type of nodes, namely, 1) the controller/verification node, which maintains the updated flow rules table information and 2) the request/response node, which updates its flow rules table in a blockchain network.

\subsection{Fog computing}
Fog computing, also called edge computing, is a highly virtualized platform that enables computing and storage between end-users and the data Center of the traditional cloud computing \cite{51}. Without the third parties, Fog devices can communicate with each other. However, the blockchain technique can be used to facilitate communications between fog nodes and IoT devices. Huang et al. \cite{46} proposed a fair payment scheme for outsourcing computations of Fog devices. Based on the bitcoin, this scheme considers the following security properties, namely, completeness, fairness, and accountability. 

\subsection{Distributed P2P Applications}
In distributed peer-to-peer (P2P) applications for the IoT, the IoT devices self-organize and cooperate for a new breed of applications such as collaborative movies, forwarding files, delivering messages, electronic commerce, and uploading data using sensor networks. To incentivize users for cooperation, He et al. \cite{24} proposed a truthful incentive mechanism based on the blockchain technique for dynamic and distributed P2P environments. To prevent selfish users and defend against the collusion attacks, this scheme proposed a pricing strategy, which allows intermediate nodes to obtain rewards from blockchain transactions due to their contribution to a successful delivery.

\subsection{Crowdsensing Applications}
The emerging mobile crowdsensing paradigm is a novel class of mobile IoT applications (e.g., geographical sensing applications). Wang et al. \cite{27} is an interesting incentive mechanism for privacy-preserving in crowdsensing applications based on the blockchain cryptocurrencies. Specifically, this mechanism can eliminate the security and privacy issues using the miners' verifiable data quality evaluation to deal with the impersonation attacks in the open and transparent blockchain. In addition, to achieve k-anonymity privacy protection, the mechanism uses a node cooperation method for participating users.

\subsection{Data storage}
The data storage can deal with heterogeneous data resources for IoT-based data storage systems. How to share and protect these sensitive data are the main challenges in IoT data storage. Based on the blockchain technology, Jiang et al.~\cite{41} proposed a private keyword search, named Searchain, for decentralized storage. The Searchain architecture includes two component, namely, 1) transaction nodes in a peer-to-peer structure and 2) a blockchain of all the ordered blocks. In addition, the Searchain architecture can provide user privacy, indistinguishability, and accountability.

\begin{table*}[!t]
	\centering
	\caption{Existing Research on Anonymity and Privacy for Bitcoin Systems}
	\label{Table:Tab1}
	\setlength{\tabcolsep}{2pt}
		\vspace*{-\baselineskip}
		\renewcommand{\arraystretch}{1.5}
{\scriptsize
\begin{tabular}{p{0.3in}|p{1.2in}|p{3in}|p{1.9in}} \toprule
\rowcolor{black!15}\textbf{Year} & \textbf{Protocol} & \textbf{Countermeasures} & \textbf{Security models} \\ \midrule
2013 & CoinSwap \cite{61} & - The protocol requires four published transactions & - Anonymity \\ \hline 
\rowcolor{black!5} 2013 & CoinJoin \cite{65} & - Each user check the mixing transaction before signing on it & - Anonymity \\ \hline 
2013 & ZeroCoin \cite{68} & - Decentralized e-cash scheme with a tuple of randomized algorithms (Setup, Mint, Spend, Verify)\newline - RSA accumulators and non-interactive zero-knowledge signatures of knowledge & - Anonymity \\ \hline 
\rowcolor{black!5} 2014 & Mixcoin \cite{59} & - Cryptographic accountability\newline - Randomized mixing fees & - Anonymity \\ \hline 
2014 & Xim \cite{64} & - Anonymous decentralized pairing & - Anonymity \\ \hline 
\rowcolor{black!5} 2014 & CoinShuffle \cite{67} & - Requires only standard cryptographic primitives & - Anonymity \\ \hline 
2014 & Zerocash \cite{69} & - Publicly-verifiable preprocessing zero-knowledge & - Privacy-preserving \\ \hline 
\rowcolor{black!5} 2015 & Blindcoin \cite{60} & - Blind signature scheme & - Anonymity \\ \hline 
2015 & CoinParty \cite{66} & - Combination of decryption mixnets with threshold signatures & - Anonymity \\ \hline 
\rowcolor{black!5} 2016 & Blindly Signed Contract \cite{62} & - Blind signature scheme & - Anonymity \\ \hline 
2017 & TumbleBit \cite{63} & - Replaces on-blockchain payments with off-blockchain puzzle solving & - Anonymity \\ \hline 
\rowcolor{black!5} 2017 & Kanemura et al. \cite{37} & - The privacy metric "Deniability" & - Privacy-preserving \\ \hline 
2017 & Wang et al. \cite{42} & - Elliptic curve cryptography & - Privacy-preserving \\ \hline 
\rowcolor{black!5} 2017 & Wang et al. \cite{45} & - Homomorphic paillier encryption system & - Privacy-preserving \\ \hline 
2018 & Liu et al. \cite{55} & - Ring signature\newline - Elliptic curve digital signature algorithm & - Privacy-preserving\newline - Anonymity \\ \hline 
\rowcolor{black!5} 2018 & Huang et al. \cite{46} & - Commitment-based sampling scheme & - Security requirement of completeness\newline - Security requirement of fairness\newline - Security requirement of accountability \\ \bottomrule
\end{tabular}
}
\end{table*}

\subsection{Bitcoin}
Launched in 2009, Bitcoin is the peer-to-peer (P2P) payment network that does not need any central authorities. Based on the core technique of blockchain, Bitcoin users do not use real names; instead, pseudonyms are used. Therefore, Bitcoin is based on three main technical components: transactions, consensus Protocol, and communication network. 

The existing research on anonymity and privacy for Bitcoin system are presented in Tab. \ref{Table:Tab1}. Khalilov and Levi \cite{34} have published an interesting investigation on anonymity and privacy in Bitcoin-like digital cash systems. Specifically, the study classified the methods of analyzing anonymity and privacy in Bitcoin into four categories, namely, 1) Transacting, 2) Utilizing off-network information, 3) Utilizing network, and 4) Analyzing blockchain data.

As discussed by Wang et al. \cite{42}, Bitcoin works in practice, but not in theory,and the main issue is how to protect the potential buyers' privacy in Bitcoin using the public key infrastructure. Wang et al. \cite{42} studied the designated-verifier proof of assets for bitcoin exchange using elliptic curve cryptography. Specifically, the authors proposed a privacy-preserving scheme, named DV-PoA, which can satisfy unforgeability. Note that the DV-PoA scheme uses elliptic curve discrete logarithm problem, elliptic curve computational Diffie-Hellman problem, and collision-resistance of cryptographic hash function. In addition,  to protect the privacy of simplified payment verification (SPV) clients, Kanemura et al. \cite{37} proposed a privacy-preserving Bloom filter design for an SPV client based on $\gamma -$Deniability.

By removing the trusted third party, Qin et al. \cite{44} proposed a distributively blockchain-based PKI for Bitcoin system, named Cecoin. To ensure the consistency, Cecoin uses an incentive mechanism and a distributed consensus protocol. To provide multi-certificate services and identity assignment,  Cecoin converts a triple (address, domain, cert) to a tuple (key, address, cert), and key represents path of cert in the tree. Therefore, to protect the transaction privacy in Bitcoin, Wang et al. \cite{45} proposed a framework by adding the homomorphic Paillier encryption system to cover the plaintext amounts in transactions. To solve the trust problem in Bitcoin, Huang et al. \cite{46} proposed a commitment-based sampling scheme instead of ringer, which can be used for generic computations in the outsourcing computing.

\section{Threat models for Blockchain}\label{sec:3}

\begin {figure*}[!ht]
\centering
\begin{adjustbox}{width=0.9\textwidth}
\begin{tikzpicture}[
Level1/.style   = {rectangle, draw, rounded corners=1pt, thin, align=center, fill=blue!8},
Level2/.style = {rectangle, draw, rounded corners=1pt, thin, align=center, fill=black!8, text width=8em},
Level3/.style = {rectangle, draw, rounded corners=1pt, thin, align=left, fill=green!2, 	text width=7.8em},
level 1/.style={sibling distance=35mm},
edge from parent path={(\tikzparentnode.south)
	-- +(0,-8pt)
	-| (\tikzchildnode)}]

\node[Level1] {Threat models}
child {node[Level2] (c1) {Identity-based attacks}}
child {node[Level2] (c2) {Manipulation-based attacks}}
child {node[Level2] (c3) {Cryptanalytic attacks}}
child {node[Level2] (c4) {Reputation-based attacks}}
child {node[Level2] (c5) {Service-based attacks}};

\begin{scope}[every node/.style={Level3}, node distance=3mm]
\node [below = of c1, xshift=0.5em] (c11) {Key attack~\cite{3}};
\node [below = of c11] (c12) {Replay attack~\cite{57,3}};
\node [below = of c12] (c13) {Impersonation attack~\cite{23, 57,3}};
\node [below = of c13] (c14) {Sybil attack~\cite{40}};

\node [below = of c2, xshift=0.5em] (c21) {False data injection attack~\cite{25}};
\node [below = of c21] (c22) {Tampering attack~\cite{45}};
\node [below= of c22] (c23) {Overlay attack~\cite{45}};
\node [below = of c23] (c24) {Modification attack~\cite{57,3}};
\node [below = of c24] (c25) {Man-in-the-middle attack~\cite{57,3}};

\node [below = of c3, xshift=0.5em] (c31) {Quantum attack~\cite{11}};

\node [below = of c4, xshift=0.5em] (c41) {Hiding Blocks attack~\cite{40}};
\node [below = of c41] (c42) {Whitewashing~\cite{40}};

\node [below = of c5, xshift=0.5em] (c51) {DDoS/DoS attack~\cite{22,43,55, 57}};
\node [below = of c51] (c52) {Refusal to Sign attack~\cite{40}};
\node [below = of c52] (c53) {Double-spending attack~\cite{45, 33}};
\node [below = of c53] (c54) {Collusion attack~\cite{24}};
\end{scope}

\foreach \value in {1,2,3,4}
\draw[-] (c1.195) |- (c1\value.west);

\foreach \value in {1,...,5}
\draw[-] (c2.195) |- (c2\value.west);

\foreach \value in {1}
\draw[-] (c3.195) |- (c3\value.west);

\foreach \value in {1,2}
\draw[-] (c4.195) |- (c4\value.west);

\foreach \value in {1,...,4}
\draw[-] (c5.195) |- (c5\value.west);
\end{tikzpicture}

\end{adjustbox}
\caption{Classification of threat models for Blockchain.}
\label{fig_threats}
\end{figure*}
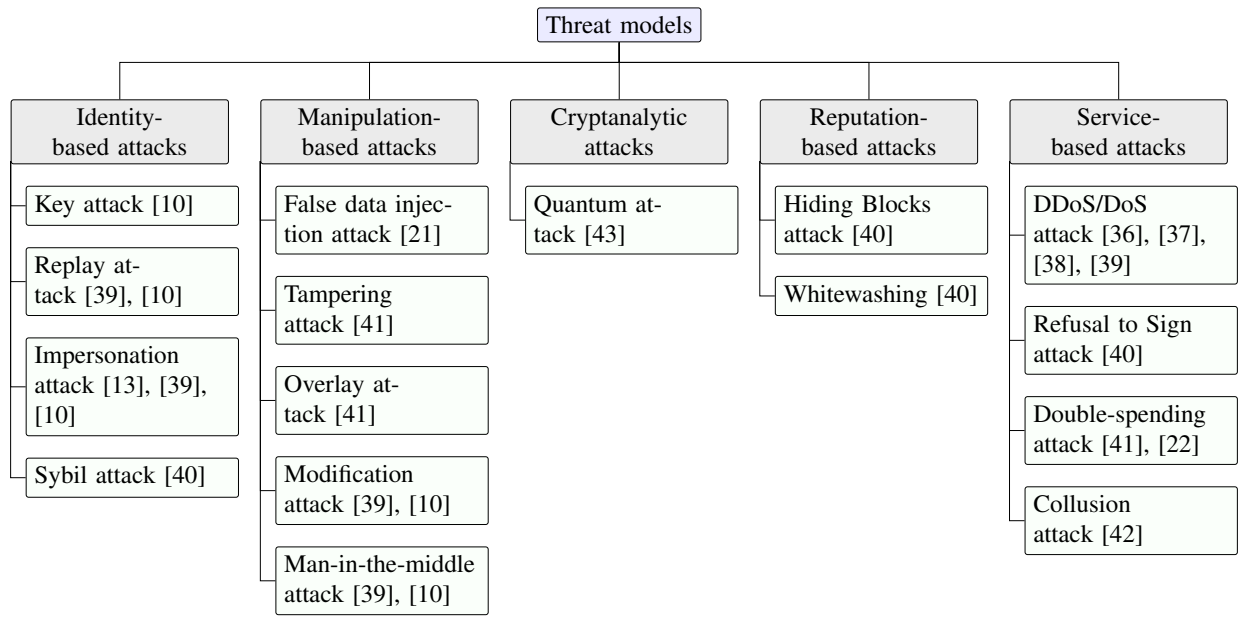

In this section, we present and describe the threat models that are considered by the blockchain protocols in IoT networks. A summary of 16 attacks are given in Table \ref{Table:Tab3}, and are classified into the following five main categories: identity-based attacks, manipulation-based attacks, cryptanalytic attacks, reputation-based attacks, and service-based attacks, as presented in Figure \ref{fig_threats}.

\subsection{Identity-based attacks}
The attacks under this category forge identities to masquerade
as authorized users, in order to get access to the system and manipulate it. We classify four attacks, namely: Key attack,  Replay attack, Impersonation attack, and Sybil attack.

\begin{itemize}

\item \textit{Key attack}: This attack is defined in the context of a system combining electric vehicles and charging piles, as follows: "If the private key of an electric vehicle that has been used for longtime leaks, the attacker can impersonate this electric vehicle to deceive others" \cite{3}. To deal with this attack, LNSC protocol \cite{3} provides a mutual authentication mechanism between the electric vehicles and charging piles. To this end, it employs the elliptic curve encryption to calculate the hash functions, and hence it ensures resiliency against the key leakage attack.

\item \textit{Replay attack}: The aim of this attack is to spoof the identities of two parties, intercept their data packets, and relay them to their destinations without modification. To resist against this attack, LNSC \cite{3} uses the idea of elliptic curve encryption to calculate the hash functions. On the other hand, BSein \cite{57} uses a fresh one-time public/private key pair, which is generated for each request, to encrypt the message and compute the Message Authentication Code (MAC). In this way, the replay attack can be detected.

\item \textit{Impersonation attack}: An adversary tries to masquerade
as a legitimate user to perform unauthorized operations. As presented in Table \ref{Table:Tab3}, there are three methods that are proposed to protect against this attack. The idea of elliptic curve encryption to calculate the hash functions, is proposed by LNSC protocol \cite{3}. Wang et al. \cite{27} propose a distributed incentive-based cooperation mechanism, which  protects the user's privacy as well as a transaction verification method of the node cooperation. The mechanism hides the user's privacy information within a group, and ensures their protection from the impersonation attack. BSeIn \cite{57}, on the other hand, uses the idea of attribute-based signatures, i.e., only legitimate terminals can generate a valid signature, and hence any impersonation attempt will be detected when its corresponding authentication operation fails.

\item \textit{Sybil attack}: Under this attack, an adversary creates many fake identities. By performing many interactions in the network, the adversary can gain a large influence within the community, i.e,  increasing/decreasing the reputation of some agents. TrustChain \cite{40} addresses this issue by creating an immutable chain of temporally ordered interactions for each agent. It computes the trustworthiness of agents in an online community with Sybil-resistance by using prior transactions as input. It ensures that agents who use resources from the community also contribute back.

\end{itemize}

\subsection {Manipulation-based attacks} They involve an unauthorized access and tamper of data. In this category, four attacks are classified, namely:  False data injection attack, Tampering attack, Overlay attack, and Modification attack

\begin{itemize}

\item \textit{False data injection attack}: The aim of this attack is to compromise the data integrity of the control system to make it take wrong control decisions. Liang et al. \cite{25} considers the meter node as a private blockchain network. In addition, the interactions among the nodes are based on a consensus mechanism, which consists in  executing  a distributed voting algorithm. Each node can verify the integrity of the received data. The latter is considered correct when a positive agreement is reached. 

\item \textit{Tampering attack}: The adversary may tamper the bitcoin
transactions of the bitcoin addresses, amounts and other information after signing. To prevent this attack, Wang et al. \cite{45} use a public-key cryptosystem that is compatible with the existing Bitcoin system. They propose adding the homomorphic Paillier encryption system to cover the plaintext amounts in transactions, and the encrypted amounts will be checked by the Commitment Proof.

\item \textit{Overlay attack}: It means that the attacker adds a forgery encrypted amount to the original encrypted amount under the receiver's public key. In \cite{45}, this attack is detected as every transaction
is embedded with an timestamp to mark its uniqueness.
Different inputs under the same trader can be distinguished and linked to
the different transactions, and hence resistance against the overlay attack is ensured.

\item \textit{Modification attack}: It consists in modifying the broadcast transaction or the response message. To deal with this attack, LNSC \cite{3} uses the idea of elliptic curve encryption to calculate the hash functions. BSeIN \cite{57}, on the other hand, employs the attribute signature and the MAC.

\item \textit{Man-in-the-middle attack}: An attacker by spoofing
the identities of two parties can secretly relay and even modify the communication between these parties, which believe they are communicating directly, but in fact the whole conversation is under the control of the attacker. To resist against this attack, BSeIn \cite{57} provides secure mutual authentication. In \cite{3}, LNSC provides mutual authentication by using elliptic curve encryption to calculate the hash functions.

\end{itemize}

\subsection{Cryptanalytic attacks} They aim to break the cryptographic algorithm and expose its keys. In \cite{11} the quantum attack is investigated in blockchain. This attack is designed to solve the elliptic  curve digital logarithm, i.e., derive the private key from the elliptic curve public key. In this way, an adversary can sign unauthorized transactions and forge the valid signature of users. To deal with this issue, Yin et al. \cite{11} uses the idea of lattice-based signature scheme., which allows deriving many sub-private keys from the seed in the deterministic wallet of blockchain.

\subsection{Reputation-based attacks} An agent manipulates his reputation by changing it to a positive one. In this category, we can find the following attacks, namely: Hiding Blocks attack, and Whitewashing attack.

\begin{itemize}

\item \textit{Hiding Blocks attack}: Under this attack, an agent only exposes transactions that have a positive impact on his reputation and
hides the ones with negative reputation. In \cite{40}, an immutable chain of temporally ordered interactions for each agent. Since each record has a sequence number, any agent in the network can request specific records of others. The requested agents cannot refuse to provide their records. Otherwise, other agents will stop interacting with them.

\item \textit{Whitewashing}: When an agent has negative reputation, it can get rid of its identity and make a new one. There is no way to prevent this behavior. However, it is suggested in \cite{40} to give lower priorities to the agents of new identities when applying the allocation policy.

\end{itemize}

\subsection{Service-based attacks} They aim either to make the service unavailable or make it behave differently from its specifications. Under this category, we can find the following attacks:

\begin{itemize}

\item \textit{DDoS/DoS attack}: It involves sending a large amount of requests to cause the failure of the blockchain system. As shown in Table \ref{Table:Tab3}, there are four methods that are proposed to deal with this attack. The idea of distributed SDN architecture is proposed by DistBlockNet protocol in \cite{22}. CoinParty \cite{43} proposes the idea of decentralized mixing service. Liu et al. \cite{55} employ a ring-based signature with Elliptic Curve Digital Signature Algorithm (ECDSA). The resilience against DOS in BSeIn \cite{57} is achieved by limiting the block size, checking the maximum number of attribute signatures for the transaction input, and using multi-receivers encryption to provide confidentiality for authorized participants. 

\item \textit{Refusal to Sign attack}: A malicious agent can decide to not sign a transaction that is not in his favor. Although preventing this attack is not possible, punishment measures can be taken against the refusal agents. It is proposed in \cite{40} to not interact with the malicious agent, or split the transactions in smaller amounts. If an agent refuses to sign a transaction, the interaction is aborted.

\item \textit{Double-spending attack}: It means that the attackers
spend the same bitcoin twice to acquire extra amounts. In \cite{45}, the Time-Stamp and the Proof-of-Work mechanism is used. In \cite{33}, a multi-signature transaction is employed, where a minimum number of keys must sign a transaction before spending tokens.

\item \textit{Collusion Attack}: Nodes can collude with each other and behave selfishly to maximize their profit.  In \cite{24}, an incentive mechanism and pricing strategy is proposed to thwart the selfish behaviors.

\end{itemize}

\begin{table*}[!t]
	\centering
	\caption{Existing research for blockchain-based IoT security and privacy}
	\label{Table:Tab2}
	\setlength{\tabcolsep}{2pt}
		\vspace*{-\baselineskip}
		\renewcommand{\arraystretch}{1.5}
{\scriptsize
\begin{tabular}{p{0.3in}|p{0.4in}|p{1in}|p{0.9in}|p{1.2in}|p{2in}|p{0.8in}} \toprule
\rowcolor{black!15}\textbf{Year} & \textbf{Scheme} & \textbf{Blockchain model} & \textbf{Security model} & \textbf{Goal} & \textbf{Performance (+) and limitation (-)} & \textbf{Comp. complexity} \\ \midrule
2016 & Aitzhan and Svetinovic \cite{33} & - Blockchain technology with multi signatures and anonymous encrypted message propagation streams & - Privacy preserving & - Enables peers to anonymously negotiate energy prices and securely perform trading transactions & + Combat double-spending attacks\newline - A formal proof is not provided on the Sybil-resistance\newline  & Medium \\ \hline 
\rowcolor{black!5} 2017 & Otte et al. \cite{40} & - Every participant grows and maintains their own chain of transactions & - Distributed trust & - Providing strict bounds on the profitability of a Sybil attack & + A formal proof is provided on the Sybil-resistance\newline - Authentication is not considered  & Up to $2n+1$ max-flow computations \\ \hline
2017 & Kanemura el al. \cite{37} & - Blockchain technology with Deniability & - Privacy preserving & - Improving the privacy level of a simplified payment verification client & + True positive Bitcoin addresses are hidden by the false positives in a Bloom filter\newline - Authentication is not considered & Medium \\ \hline
\rowcolor{black!5} 2017 & Wang et al. \cite{45} & - Blockchain technology with the Paillier cryptosystem for encryption and decryption & - Preserving transaction privacy & - Achieving delicate anonymity and prevents active and passive attacks & + Robust transaction\newline + Prevent the following attacks: Tampering attack, Overlay attack, Double-spending attack\newline - Sybil-resistance & $T_{dec}=2T_m+2iT_E$ \\ \hline 
2018 & Yin et al. \cite{11} & - Quantum attack in the blockchain & - Transaction authentication & - Resisting quantum attack, while maintaining the wallet lightweight & + Strongly unforgeable under chosen message attack\newline - The Sybil-resistance is not considered  & The length of signature is $O(1)$ \\ \hline 
\rowcolor{black!5} 2018 & Jong-Hyouk Lee \cite{12} & - Consortium Blockchain & - Identity and authentication management & - Creating a new ID as a Service & + It can be implemented as a cloud platform\newline - The threat model is not defined & Medium \\ \hline 
2018 & Fan et al. \cite{2} & - The blockchain is a public, tamper-resistant ledger & - Privacy preserving\newline - Access control & - Achieve the goal of every data owner's complete control & + Backward security \newline + Forward security\newline - The Sybil-resistance is not considered & $M+\ T_m$ \\ \hline
\rowcolor{black!5} 2018 & Wang  et al. \cite{27} & - Blockchain based incentive mechanism & - Privacy preserving & - Achieve \textit{k}-anonymity privacy protection & + Resist the impersonation attacks in the open and transparent blockchain\newline - The collusion attacks is not analysed  & Medium \\ \hline
2018 & Lin et al. \cite{29} & - ID-based linearly homomorphic signature & - Authentication & - Avoiding the shortcomings of the use of public key certificates & + Secure against existential forgery on adaptively chosen message and ID attack in the random oracle model\newline - Adaptation with the Blockchain is not analyzed & High \\ \hline 
\rowcolor{black!5} 2018 & Li et al. \cite{31} & - Blockchain based incentive mechanism & - Privacy preserving\newline - Authentication & - Achieving  privacy-preserving in forwarding announcements & + Maintains the reliability of announcements\newline + Achieve Sybil-resistance\newline - Location privacy is not considered  & Medium \\ \hline
2018 & Ziegeldorf et al. \cite{43} & - Blockchain technology with Deniability & - Anonymity\newline - Deniability & - Achieving correctness, anonymity, and deniability & + Resilience against DoS attacks from malicious attackers\newline + Compatible with other crypto-currencies which use the same ECDSA primitive, e.g., Litecoin and Mastercoin\newline - Double-spending attacks is not considered & Medium \\ \hline
\rowcolor{black!5} 2018 & Yang  et al. \cite{53} & - The blocks maintain the proofs produced by the cloud server & - Accountable traceability & - Achieving public verification without any trusted third party & + Achieve public verification\newline + Efficient in communication as well as in computation\newline - Tampering attack is not considered & The data owner conducts $(2+{log}_2m)$ hash computations \\ \hline
2018 & Hu et al. \cite{54} & - The Ethereum blockchain & - Distributed trust & - Saving on the overall deployment and operational costs & + Low-cost, accessible, reliable and secure payment scheme\newline - Accountable traceability is not considered & Low bandwidth \\ \hline
\rowcolor{black!5} 2018 & Liu et al. \cite{55} & - The blockchain based on the ring signature with elliptic curve digital signature algorithm (ECDSA) & - Preserving transaction privacy & - Help Bitcoin users protect their account and transaction information & + Resistant to DoS attacks\newline + Prevent the mixing server from mapping input transactions\newline + Anonymity and scalability\newline - Double-spending attacks is not considered & High \\ \hline
2018 & Lin et al. \cite{57} & - The structure of blocks is similar to that in Bitcoin & - Authentication\newline - Access control & - Enforce fine-grained access control polices & + Resilience to hijacking attacks, user impersonation attacks, DDoS attacks, modification attacks, replay attacks, and man-in-the-middle attacks\newline + Mutual authentication\newline + Session key agreement\newline + Perfect forward secrecy\newline - The Sybil-resistance is not considered & Medium \\ \hline
\rowcolor{black!5} 2018 & Lin et al. \cite{58} & - The Ethereum blockchain & - Authentication & - Solving the existing intractability issue in transitive signature & + Update the certificates without the need to re-sign the nodes\newline + Provide a proof when the edge between two vertices does not exist\newline - Access control is not considered compared to the scheme in \cite{57} & Signature size: 2 points in ${{\rm Z}}^*_q$ \\ \bottomrule
\end{tabular}
\begin{flushleft}
\underline{Notations :}\\
$M$: The time for one exponentiation;\\ $T_m$: The size of the ciphertext; \\$T_{dec}$: The time for decryption; \\$T_m$: The unit of modular multiplication time; \\$T_E$: The unit of modular exponentiation time
\end{flushleft}
}
\end{table*}

\section{Existing research on security and privacy in blockchain-based IoT}\label{sec:4}

Table \ref{Table:Tab2} summarizes research for blockchain-based IoT security and privacy.

\subsection{Authentication}

In \cite{58}, Lin et al. proposed a novel transitively closed undirected graph authentication scheme that can support blockchain-based identity management systems. In comparison to other competing authentication schemes, their proposal provides an additional capability of dynamically adding or deleting nodes and edges. Moreover, this novel scheme that was built on Ethereum solves the authentication problem of non-existent edges, which is a known challenge in transitive signature schemes. Lin et al.in \cite{57} proposed a novel blockchain-based framework that can ensure a secure remote user authentication. The proposed framework combines attribute-based signatures,  multi-receivers encryption and Message Authentication Code. In \cite{31}, Li et al. proposed a novel privacy-preserving Blockchain-based announcement network for Vanets that is based on a threshold authentication protocol called Echo-Announcement. 

Authors in \cite{29} proposed an ID-based linearly homomorphic signature schemes that can be used for realizing authentication in blockchains. The system allows a signer to produce linearly homomorphic signatures, and hence it avoids the shortcomings of public-key certificates. In addition, it is shown to be robust against several attacks. In \cite{12} authors introduced the concept of blockchain as a service. Their proposed blockchain based-ID as a Service (BIDaaS) mechanism, is a new type of IDaaS that can be used for identity and authentication management. Authentication can be achieved without the use of any preregistered information of the user. Finally in \cite{11} authors cope with the problem of keeping the wallet in a relatively small size while ensuring the robustness of transaction authentication by introducing a novel anti-quantum transaction authentication scheme.

\subsection{Privacy-preserving}

In the core of blockchain philosophy lies the private key that can unlock the cryptographic protection of the digital assets.  
The private key becomes the highest vulnerability of a blockchain system whether it is stored on a piece of paper, screen, disk, in local memory or in the cloud. Users tend to use digital wallets that can be either software or hardware, e.g. Trezor or Keepkey, which are vulnerable to various attacks like fault injections \cite{boireau2018securing}. 

Another solution that is gaining ground nowadays is the use of hardware security modules (HSMs), a crypto-processor that securely generates, protects and stores keys. The entire cryptographic key lifecycle happens inside the HSM. An HSM can be a standalone device that operates offline or can be embedded in a server, can be hardened against tampering or damage, and is usually located in a physically secure area to prevent unauthorized access. Finally a new generation of ultra-secure PCs that have embedded an HSM and requires two-factor  authentication is recently introduced. This PC can be protected against physical attacks with a tamper-proof casing and mechanisms like automatic erasion of the private key in case of any breach of the  embedded physical or logical security controls \cite{orwl}. Using trusted computers both as secure digital wallets and blockchain nodes. Security assurance of users and organizations need in order to trust this new technology can be provided in the near future.

To achieve k-anonymity privacy protection, Wang et al. \cite{27} use a node cooperation verification approach, in which each group contains K nodes to meet the objective of K-anonymity protection. Aitzhan et al. \cite{33} proposed an idea that protects parties from passive eavesdropping by hiding non-content data. For enhancing the transaction privacy in Bitcoin, Wang et al. \cite{45} achieve transaction by using cryptographic methods, i.e., employs the public-key system. Through the standard ring signature and ECDSA unforgeability, Liu et al. \cite{55} proposed an idea that can achieve the anonymity.

One other aspect of privacy in blockchain systems is about anonymity. Although it is possible to design an almost immutable, tamper-resistant transaction, this transaction can be seen throughout all of the nodes on the blockchain network. One promising research on supporting private transactions inside a blockchain  is zk-STARKs, which combines zCash and Ethereum. The combination of both technologies makes it possible to keep anonymity when conducting payments, blind auctions, and even voting \cite{44}.

\subsection{Trust}
A blockchain-based payment scheme that is stet up in a remote region setting was introduced in \cite {54}. The proposed scheme is assumed to have an  intermittent connectivity to a bank's central system. Distributed trust is accomplished with the use of a two-layer architecture, where the bank authorizes a set of selected villagers to act as miners who on their turn authorize transactions among villagers with tokens and the bank. In \cite{40} authors present a mechanism where every participant grows and maintains his own chain of transactions. The proposed  approach  provides distributed trust, without the need of any gatekeeper, while being robust against Sybil attacks.

\section{Open questions and research challenges}\label{sec:5}
To complete our overview, we outline both open questions and research challenges that could improve the capabilities and effectiveness of blockchain for the IoT, summarized in the following recommendations:

\subsection{Resiliency against Combined Attacks}
As presented in this survey, many security solutions for bloackchain-based IoT have been proposed in the literature, each of which is designed to tackle different security issues and threat models. The main question that might arise is how to design a security solution that can be resilient against combined attacks while taking into account the implementation feasibility of the solution, especially in case of low resource-constrained IoT devices.

\subsection{Dynamic and Adaptable Security Framework}
Heterogeneous devices are deployed in the IoT network, ranging from low-power devices to high-end servers. Hence, a single security solution cannot be deployed for all the blockchain-based IoT architectures due to the different amount of  resources that are provided. Therefore, the security solution should initially adapt itself to the existing resources, and decide which security services to offer, so as to meet the minimum security requirements of the end-users. Thus, one of the challenges that should receive more attention in the future is how to design such a dynamic and adaptable security framework for blockchain-based IoT architectures.

\subsection{Compliance with GDPR}
The new regulation that took effect from the 25th of May of 2018 grants end-users new powers over personal data, and places new obligations upon data controllers. In a purely decentralized blockchain system there are not any accountable data processors making the implementation of GDPR difficult. Also on the other hand, the right to be forgotten that is one of the main guidelines of GDPR is contradictory to the core idea behind blockchain technology. On public blockchains, all data is replicated and shared across all machines in the network and only the combination with trusted hardware solutions can solve this issue. Some initial attempts are already undertaken \cite{lind2017teechain, bentov2017tesseract} but this is a promising area of research for the near future. 

\subsection{Energy-efficient Mining}
Mining includes the execution of the blockchain consensus algorithms such as Proof-of-Work (PoW). Besides, the blockchain grows as the users store their transactions. Therefore, more powerful miners are required to handle the consensus protocols in the blockchain. Several energy efficient consensus algorithms such as Proof-of-Space~\cite{PoS2015}, Deligated Proof-of-Space~\cite{DPos} and Proof-of-Stake~\cite{ProofofStake} and mini-blockchain~\cite{FrancaMiniBlock2015, Bruce2014} to store only recent blockchain transactions are suggested. However, resource- and power-constrained IoT devices are not always capable of meeting the substantial computational and power consumption in the processing of blockchain consensus and storing of blockchains.  Therefore, the design of energy-efficient consensus protocols is one of the significant research challenges in the blockchain technologies for IoT. 

\subsection{Social Networks and Trust Management}
When we talk about security we have to take in  mind that fake news can be a  part of a cyber attack. Large-scale rumor spreading could pose severe social and economic damages to an organization or a nation \cite{ayres2016cyberterrorism} especially with the use of online social networks. Blockchains could be a means for limiting rumor spreading as presented in \cite {chen2018towards} where a blockchain-enabled social network is presented. 

\subsection{Blockchain-specific Infrastructure}
The storage-limited IoT devices might not be able to store the large-size blockchain that grows as the blocks are appended in the blockchain. Moreover, it is commonly seen that the IoT devices store the blockchain's data that are not even useful for their own transactions. Therefore, blockchain-specific equipment that supports the decentralized storage of large-size blockchain become a challenging issue. Moreover, the address management and underlying communication protocols play significant roles in blockchain infrastructure. Besides, trustworthiness among the computational resource-enriched devices has to be established in the blockchain infrastructure. Besides, the Application Programming Interface (APIs) should be user-friendly as much as possible.

\subsection{Vehicular Cloud Advertisement Dissemination}
As presented in this survey, based on a decentralized blockchain structure, various anonymity schemes are proposed to hide the real identities in IoV. Therefore, since the vehicle's real identity, vehicle's real location, and transaction could possibly be disclosed in vehicular cloud advertisement dissemination~\cite{kong2018achieving}, critical security issues arise as follows:

\begin{itemize}
\item  How to design a single-attribute access control protocol based on blockchain technology for preserving transaction privacy in vehicular cloud advertisement dissemination?
\item  How to devise a privacy-preserving secret sharing scheme based on blockchain technology to acknowledge participation of selected vehicles in transactions? e.g., by using the homomorphic Paillier encryption system.
\item  How to design a low complexity-based authentication using the blockchain technology between RSUs and participating vehicles during the advertisement dissemination process?
\end{itemize}

\subsection{Skyline Query Processing}
Skyline query has become an important issue in database research, e.g., centralized database, distributed database, and similarity search. The surveyed schemes have not yet studied the possibility of using the skyline query with blockchain. Recently, Hua et al.~\cite{hua2018cinema} proposed a privacy-preserving online medical primary diagnosis framework, named CINEMA, which uses the skyline query. Specifically,  CINEMA framework can protect users' medical data privacy and ensure the confidentiality of diagnosis model based on a skyline diagnosis model. Therefore, how to handle the security and privacy issues when a skyline diagnosis model is constructed by a lot of blockchains? Hence, the privacy-preserving schemes based on the blockchain with skyline query are major challenges and should be investigated in the future.

\section{Conclusion}\label{sec:6}
In this paper, we surveyed the state-of-the-art of existing blockchain protocols designed for  Internet of Things (IoT) networks. We provided an overview of the application domains of blockchain technologies in IoT, e.g., Internet of Vehicles, Internet of Energy, Internet of Cloud, and Fog computing. Through extensive research and analysis that was conducted, we were able to classify the threat models that are considered by the blockchain protocols in IoT networks, into five main categories, namely, identity-based attacks, manipulation-based attacks, cryptanalytic attacks, reputation-based attacks, and service-based attacks. There still exist several challenging research areas, such as resiliency against combined attacks, dynamic and adaptable security framework, compliance with GDPR,  energy-efficient mining, social networks and trust management, blockchain-specific infrastructure, vehicular cloud advertisement dissemination, and Skyline query processing, which should be further investigated in the near future.
\bibliographystyle{IEEEtran}
\bibliography{bare_jrnl}

\begin{thebibliography}{10}
\providecommand{\url}[1]{#1}
\csname url@samestyle\endcsname
\providecommand{\newblock}{\relax}
\providecommand{\bibinfo}[2]{#2}
\providecommand{\BIBentrySTDinterwordspacing}{\spaceskip=0pt\relax}
\providecommand{\BIBentryALTinterwordstretchfactor}{4}
\providecommand{\BIBentryALTinterwordspacing}{\spaceskip=\fontdimen2\font plus
\BIBentryALTinterwordstretchfactor\fontdimen3\font minus
  \fontdimen4\font\relax}
\providecommand{\BIBforeignlanguage}[2]{{%
\expandafter\ifx\csname l@#1\endcsname\relax
\typeout{** WARNING: IEEEtran.bst: No hyphenation pattern has been}%
\typeout{** loaded for the language `#1'. Using the pattern for}%
\typeout{** the default language instead.}%
\else
\language=\csname l@#1\endcsname
\fi
#2}}
\providecommand{\BIBdecl}{\relax}
\BIBdecl

\bibitem{IDCIoT2}
``{IDC, Worldwide Internet of Things Forecast, 2015--2020},'' {IDC} \#256397.

\bibitem{IDCIoT}
``{IDC, Worldwide Internet of Things Forecast Update 2015--2019},'' Feb. 2016,
  {Doc} \#US40983216.

\bibitem{Miorandi2012internet}
D.~Miorandi, S.~Sicari, F.~De~Pellegrini, and I.~Chlamtac, ``Internet of
  things: Vision, applications and research challenges,'' \emph{Ad Hoc Netw.},
  vol.~10, no.~7, pp. 1497--1516, Sept. 2012.

\bibitem{Puthal2018CM}
D.~Puthal, N.~Malik, S.~P. Mohanty, E.~Kougianos, and G.~Das, ``Everything you
  wanted to know about the blockchain: Its promise, components, processes, and
  problems,'' \emph{{IEEE} Consumer Electronics Mag.}, vol.~7, no.~4, pp.
  6--14, July 2018.

\bibitem{51}
M.~Mukherjee, R.~Matam, L.~Shu, L.~Maglaras, M.~A. Ferrag, N.~Choudhury, and
  V.~Kumar, ``Security and privacy in fog computing: Challenges,'' \emph{IEEE
  Access}, vol.~5, pp. 19\,293--19\,304, 2017.

\bibitem{SwanBlockchainBook}
M.~Swan, \emph{Blockchain: blueprint for a new economy}, 1st~ed.\hskip 1em plus
  0.5em minus 0.4em\relax \'{O}Reilly Media, Jan. 2015.

\bibitem{Tschorsch2016}
F.~Tschorsch and B.~Scheuermann, ``Bitcoin and beyond: A technical survey on
  decentralized digital currencies,'' \emph{{IEEE} Commun. Surveys {\&} Tut.},
  vol.~18, no.~3, pp. 2084--2123, Mar. 2016.

\bibitem{52}
S.~Nakamoto, ``Bitcoin: A peer-to-peer electronic cash system,'' 2008.

\bibitem{Wilson2015}
D.~Wilson and G.~Ateniese, ``From pretty good to great: Enhancing {PGP} using
  bitcoin and the blockchain,'' in \emph{Network and System Security}.\hskip
  1em plus 0.5em minus 0.4em\relax Springer International Publishing, 2015, pp.
  368--375.

\bibitem{3}
X.~Huang, C.~Xu, P.~Wang, and H.~Liu, ``{LNSC: A} security model for electric
  vehicle and charging pile management based on blockchain ecosystem,''
  \emph{IEEE Access}, vol.~6, pp. 13\,565--13\,574, 2018.

\bibitem{13}
A.~Dorri, M.~Steger, S.~S. Kanhere, and R.~Jurdak, ``{BlockChain: A Distributed
  Solution to Automotive Security and Privacy},'' \emph{IEEE Commun. Mag.},
  vol.~55, no.~12, pp. 119--125, dec 2017.

\bibitem{20}
A.~Lei, H.~Cruickshank, Y.~Cao, P.~Asuquo, C.~P.~A. Ogah, and Z.~Sun,
  ``{Blockchain-Based Dynamic Key Management for Heterogeneous Intelligent
  Transportation Systems},'' \emph{IEEE Internet Things J.}, vol.~4, no.~6, pp.
  1832--1843, dec 2017.

\bibitem{23}
J.~Kang, R.~Yu, X.~Huang, S.~Maharjan, Y.~Zhang, and E.~Hossain, ``{Enabling
  Localized Peer-to-Peer Electricity Trading Among Plug-in Hybrid Electric
  Vehicles Using Consortium Blockchains},'' \emph{IEEE Trans. Ind.
  Informatics}, vol.~13, no.~6, pp. 3154--3164, dec 2017.

\bibitem{31}
L.~Li, J.~Liu, L.~Cheng, S.~Qiu, W.~Wang, X.~Zhang, and Z.~Zhang,
  ``{CreditCoin: A Privacy-Preserving Blockchain-Based Incentive Announcement
  Network for Communications of Smart Vehicles},'' \emph{IEEE Trans. Intell.
  Transp. Syst.}, pp. 1--17, 2018.

\bibitem{36}
Z.~Yang, K.~Zheng, K.~Yang, and V.~C.~M. Leung, ``{A blockchain-based
  reputation system for data credibility assessment in vehicular networks},''
  in \emph{2017 IEEE 28th Annu. Int. Symp. Pers. Indoor, Mob. Radio
  Commun.}\hskip 1em plus 0.5em minus 0.4em\relax IEEE, oct 2017, pp. 1--5.

\bibitem{27}
J.~Wang, M.~Li, Y.~He, H.~Li, K.~Xiao, and C.~Wang, ``{A Blockchain Based
  Privacy-Preserving Incentive Mechanism in Crowdsensing Applications},''
  \emph{IEEE Access}, vol.~6, pp. 17\,545--17\,556, 2018.

\bibitem{FengTian2016}
F.~Tian, ``An agri-food supply chain traceability system for {C}hina based on
  {RFID} {\&} blockchain technology,'' in \emph{Proc. IEEE 13th Int. Conf. on
  Service Systems and Service Management ({ICSSSM})}, June 2016.

\bibitem{32}
Z.~Li, J.~Kang, R.~Yu, D.~Ye, Q.~Deng, and Y.~Zhang, ``{Consortium Blockchain
  for Secure Energy Trading in Industrial Internet of Things},'' \emph{IEEE
  Trans. Ind. Informatics}, pp. 1--1, 2017.

\bibitem{Ahram2017}
T.~Ahram, A.~Sargolzaei, S.~Sargolzaei, J.~Daniels, and B.~Amaba, ``Blockchain
  technology innovations,'' in \emph{Proc. {IEEE} Technology {\&} Engineering
  Management Conference ({TEMSCON})}, June 2017.

\bibitem{4}
J.~Gao, K.~O. Asamoah, E.~B. Sifah, A.~Smahi, Q.~Xia, H.~Xia, X.~Zhang, and
  G.~Dong, ``{GridMonitoring: S}ecured sovereign blockchain based monitoring on
  smart grid,'' \emph{IEEE Access}, vol.~6, pp. 9917--9925, 2018.

\bibitem{25}
G.~Liang, S.~R. Weller, F.~Luo, J.~Zhao, and Z.~Y. Dong, ``{Distributed
  Blockchain-Based Data Protection Framework for Modern Power Systems against
  Cyber Attacks},'' \emph{IEEE Trans. Smart Grid}, pp. 1--1, 2018.

\bibitem{33}
N.~{Zhumabekuly Aitzhan} and D.~Svetinovic, ``{Security and Privacy in
  Decentralized Energy Trading through Multi-signatures, Blockchain and
  Anonymous Messaging Streams},'' \emph{IEEE Trans. Dependable Secur. Comput.},
  pp. 1--1, 2016.

\bibitem{Kshetri2017}
N.~Kshetri, ``Blockchain's roles in strengthening cybersecurity and protecting
  privacy,'' \emph{Telecommunications Policy}, vol.~41, no.~10, pp. 1027--1038,
  Nov. 2017.

\bibitem{FernandezCarames2018}
T.~M. Fern\'{a}ndez-Caram\'{e}s and P.~Fraga-Lamas, ``A review on the use of
  blockchain for the internet of things,'' \emph{{IEEE} Access}, pp. 1--23, May
  2018.

\bibitem{CryptocurrencyMarket}
\BIBentryALTinterwordspacing
``Crypto-currency market capitalizations,'' accessed on 15 June, 2018.
  [Online]. Available: \url{https://
  coinmarketcap.comhttps://coinmarketcap.com}
\BIBentrySTDinterwordspacing

\bibitem{BlockchainReport}
\BIBentryALTinterwordspacing
``Blockchain technology report to the {US} federal advisory committee on
  insurance,'' accessed on 15 June, 2018. [Online]. Available:
  \url{https://www.treasury.gov/initiatives/fio/Documents/McKinsey\_FACI\_Blockchain\_in\_Insurance.pdf}
\BIBentrySTDinterwordspacing

\bibitem{Bitcoin51attack}
\BIBentryALTinterwordspacing
L.~Bahack, ``Theoretical {B}itcoin attacks with less than half of the
  computational power,'' Dec. 2013, arXiv:1312.7013v1. [Online]. Available:
  \url{https://arxiv.org/pdf/1312.7013.pdf}
\BIBentrySTDinterwordspacing

\bibitem{Muneeb2016}
M.~Ali, J.~Nelson, R.~Shea, and M.~J. Freedman, ``Blockstack: A global naming
  and storage system secured by blockchains,'' in \emph{Proc. Annual Technical
  Conference ({USENIX} {ATC})}, June 2016, pp. 181--194.

\bibitem{Sankar2017}
L.~S. Sankar, M.~Sindhu, and M.~Sethumadhavan, ``Survey of consensus protocols
  on blockchain applications,'' in \emph{Proc. {IEEE} 4th Int. Conf. on
  Advanced Comput. and Commun. Syst. ({ICACCS})}, Jan. 2017.

\bibitem{Kaushik2017}
A.~Kaushik, A.~Choudhary, C.~Ektare, D.~Thomas, and S.~Akram,
  ``Blockchain-literature survey,'' in \emph{Proc. {IEEE} 2nd Int. Conf. Recent
  Trends in Electronics, Information {\&} Communication Technology ({RTEICT})},
  May 2017.

\bibitem{Khalilov2018}
M.~C.~K. Khalilov and A.~Levi, ``A survey on anonymity and privacy in
  {B}itcoin-like digital cash systems,'' \emph{{IEEE} Commun. Surveys {\&}
  Tut.}, pp. 1--1, Mar. 2018.

\bibitem{Conti2018}
M.~Conti, S.~K. E, C.~Lal, and S.~Ruj, ``A survey on security and privacy
  issues of {B}itcoin,'' \emph{{IEEE} Commun. Surveys {\&} Tut.}, pp. 1--39,
  May 2018.

\bibitem{1}
C.~Esposito, A.~{De Santis}, G.~Tortora, H.~Chang, and K.-K.~R. Choo,
  ``{Blockchain: A Panacea for Healthcare Cloud-Based Data Security and
  Privacy?}'' \emph{IEEE Cloud Comput.}, vol.~5, no.~1, pp. 31--37, Jan. 2018.

\bibitem{7}
R.~Guo, H.~Shi, Q.~Zhao, and D.~Zheng, ``{Secure Attribute-Based Signature
  Scheme With Multiple Authorities for Blockchain in Electronic Health Records
  Systems},'' \emph{IEEE Access}, vol.~6, pp. 11\,676--11\,686, 2018.

\bibitem{35}
X.~Liang, J.~Zhao, S.~Shetty, J.~Liu, and D.~Li, ``{Integrating blockchain for
  data sharing and collaboration in mobile healthcare applications},'' in
  \emph{2017 IEEE 28th Annu. Int. Symp. Pers. Indoor, Mob. Radio Commun.}\hskip
  1em plus 0.5em minus 0.4em\relax IEEE, oct 2017, pp. 1--5.

\bibitem{22}
P.~K. Sharma, S.~Singh, Y.-S. Jeong, and J.~H. Park, ``{DistBlockNet: A
  Distributed Blockchains-Based Secure SDN Architecture for IoT Networks},''
  \emph{IEEE Commun. Mag.}, vol.~55, no.~9, pp. 78--85, 2017.

\bibitem{43}
J.~H. Ziegeldorf, R.~Matzutt, M.~Henze, F.~Grossmann, and K.~Wehrle, ``{Secure
  and anonymous decentralized Bitcoin mixing},'' \emph{Futur. Gener. Comput.
  Syst.}, vol.~80, pp. 448--466, mar 2018.

\bibitem{55}
Y.~Liu, X.~Liu, C.~Tang, J.~Wang, and L.~Zhang, ``{Unlinkable Coin Mixing
  Scheme for Transaction Privacy Enhancement of Bitcoin},'' \emph{IEEE Access},
  vol.~6, pp. 23\,261--23\,270, 2018.

\bibitem{57}
C.~Lin, D.~He, X.~Huang, K.-K.~R. Choo, and A.~V. Vasilakos, ``Bsein: A
  blockchain-based secure mutual authentication with fine-grained access
  control system for industry 4.0,'' \emph{J. Netw. Comput. Appl.}, vol. 116,
  pp. 42--52, 2018.

\bibitem{40}
P.~Otte, M.~de~Vos, and J.~Pouwelse, ``{TrustChain: A Sybil-resistant scalable
  blockchain},'' \emph{Futur. Gener. Comput. Syst.}, sep 2017.

\bibitem{45}
Q.~Wang, B.~Qin, J.~Hu, and F.~Xiao, ``{Preserving transaction privacy in
  bitcoin},'' \emph{Futur. Gener. Comput. Syst.}, sep 2017.

\bibitem{24}
Y.~He, H.~Li, X.~Cheng, Y.~Liu, C.~Yang, and L.~Sun, ``{A Blockchain based
  Truthful Incentive Mechanism for Distributed P2P Applications},'' \emph{IEEE
  Access}, pp. 1--1, 2018.

\bibitem{11}
W.~Yin, Q.~Wen, W.~Li, H.~Zhang, and Z.~Jin, ``{An Anti-Quantum Transaction
  Authentication Approach in Blockchain},'' \emph{IEEE Access}, vol.~6, pp.
  5393--5401, 2018.

\bibitem{48}
M.~A. Ferrag, L.~Maglaras, A.~Argyriou, D.~Kosmanos, and H.~Janicke,
  ``{Security for 4G and 5G cellular networks: A survey of existing
  authentication and privacy-preserving schemes},'' \emph{J. Netw. Comput.
  Appl.}, vol. 101, pp. 55--82, jan 2018.

\bibitem{2}
K.~Fan, Y.~Ren, Y.~Wang, H.~Li, and Y.~Yang, ``{Blockchain-based efficient
  privacy preserving and data sharing scheme of content-centric network in
  5G},'' \emph{IET Commun.}, vol.~12, no.~5, pp. 527--532, Mar. 2018.

\bibitem{5}
J.~Gu, B.~Sun, X.~Du, J.~Wang, Y.~Zhuang, and Z.~Wang, ``{Consortium
  Blockchain-Based Malware Detection in Mobile Devices},'' \emph{IEEE Access},
  vol.~6, pp. 12\,118--12\,128, 2018.

\bibitem{47}
B.~Lee and J.-H. Lee, ``{Blockchain-based secure firmware update for embedded
  devices in an Internet of Things environment},'' \emph{J. Supercomput.},
  vol.~73, no.~3, pp. 1152--1167, mar 2017.

\bibitem{8}
O.~Novo, ``{Blockchain Meets IoT: An Architecture for Scalable Access
  Management in IoT},'' \emph{IEEE Internet Things J.}, vol.~5, no.~2, pp.
  1184--1195, Apr. 2018.

\bibitem{17}
N.~Herbaut and N.~Negru, ``{A Model for Collaborative Blockchain-Based Video
  Delivery Relying on Advanced Network Services Chains},'' \emph{IEEE Commun.
  Mag.}, vol.~55, no.~9, pp. 70--76, 2017.

\bibitem{10}
C.~Xu, K.~Wang, and M.~Guo, ``{Intelligent Resource Management in
  Blockchain-Based Cloud Datacenters},'' \emph{IEEE Cloud Comput.}, vol.~4,
  no.~6, pp. 50--59, nov 2017.

\bibitem{14}
P.~K. Sharma, M.-Y. Chen, and J.~H. Park, ``{A Software Defined Fog Node Based
  Distributed Blockchain Cloud Architecture for IoT},'' \emph{IEEE Access},
  vol.~6, pp. 115--124, 2018.

\bibitem{21}
Q.~Xia, E.~B. Sifah, K.~O. Asamoah, J.~Gao, X.~Du, and M.~Guizani, ``{MeDShare:
  Trust-Less Medical Data Sharing Among Cloud Service Providers via
  Blockchain},'' \emph{IEEE Access}, vol.~5, pp. 14\,757--14\,767, 2017.

\bibitem{49}
N.~Alexopoulos, E.~Vasilomanolakis, N.~R. Ivanko, and M.~Muhlhauser, ``Towards
  blockchain-based collaborative intrusion detection systems,'' in \emph{Proc.
  Int. Conf. Critical Inf. Infrastruct. Secur}, 2017, pp. 1--12.

\bibitem{9}
W.~Meng, E.~W. Tischhauser, Q.~Wang, Y.~Wang, and J.~Han, ``{When Intrusion
  Detection Meets Blockchain Technology: A Review},'' \emph{IEEE Access},
  vol.~6, pp. 10\,179--10\,188, 2018.

\bibitem{cruz2016cybersecurity}
T.~Cruz, L.~Rosa, J.~Proen{\c{c}}a, L.~Maglaras, M.~Aubigny, L.~Lev, J.~Jiang,
  and P.~Sim{\~o}es, ``A cybersecurity detection framework for supervisory
  control and data acquisition systems,'' \emph{IEEE Transactions on Industrial
  Informatics}, vol.~12, no.~6, pp. 2236--2246, 2016.

\bibitem{50}
K.~Kalkan and S.~Zeadally, ``Securing internet of things (iot) with software
  defined networking (sdn),'' \emph{IEEE Commun. Mag.}, 2017.

\bibitem{46}
H.~Huang, X.~Chen, Q.~Wu, X.~Huang, and J.~Shen, ``{Bitcoin-based fair payments
  for outsourcing computations of fog devices},'' \emph{Futur. Gener. Comput.
  Syst.}, vol.~78, pp. 850--858, jan 2018.

\bibitem{41}
P.~Jiang, F.~Guo, K.~Liang, J.~Lai, and Q.~Wen, ``{Searchain: Blockchain-based
  private keyword search in decentralized storage},'' \emph{Futur. Gener.
  Comput. Syst.}, sep 2017.

\bibitem{61}
\BIBentryALTinterwordspacing
G.~Maxwll, ``{Coinswap},'' 2013. [Online]. Available:
  \url{https://bitcointalk.org/index.php?topic=321228}
\BIBentrySTDinterwordspacing

\bibitem{65}
G.~Maxwell, ``Coinjoin: Bitcoin privacy for the real world,'' in \emph{Post on
  Bitcoin forum}, 2013.

\bibitem{68}
I.~Miers, C.~Garman, M.~Green, and A.~D. Rubin, ``{Zerocoin: Anonymous
  Distributed E-Cash from Bitcoin},'' in \emph{2013 IEEE Symp. Secur.
  Priv.}\hskip 1em plus 0.5em minus 0.4em\relax IEEE, may 2013, pp. 397--411.

\bibitem{59}
J.~Bonneau, A.~Narayanan, A.~Miller, J.~Clark, J.~A. Kroll, and E.~W. Felten,
  ``{Mixcoin: Anonymity for Bitcoin with Accountable Mixes},'' in \emph{Int.
  Conf. Financ. Cryptogr. Data Secur.}\hskip 1em plus 0.5em minus 0.4em\relax
  Springer Berlin Heidelberg, 2014, pp. 486--504.

\bibitem{64}
G.~Bissias, A.~P. Ozisik, B.~N. Levine, and M.~Liberatore, ``{Sybil-Resistant
  Mixing for Bitcoin},'' in \emph{Proc. 13th Work. Priv. Electron. Soc. - WPES
  '14}.\hskip 1em plus 0.5em minus 0.4em\relax New York, New York, USA: ACM
  Press, 2014, pp. 149--158.

\bibitem{67}
T.~Ruffing, P.~Moreno-Sanchez, and A.~Kate, ``{CoinShuffle: Practical
  Decentralized Coin Mixing for Bitcoin},'' in \emph{Eur. Symp. Res. Comput.
  Secur.}\hskip 1em plus 0.5em minus 0.4em\relax Springer, 2014, pp. 345--364.

\bibitem{69}
E.~B. Sasson, A.~Chiesa, C.~Garman, M.~Green, I.~Miers, E.~Tromer, and
  M.~Virza, ``{Zerocash: Decentralized Anonymous Payments from Bitcoin},'' in
  \emph{2014 IEEE Symp. Secur. Priv.}\hskip 1em plus 0.5em minus 0.4em\relax
  IEEE, may 2014, pp. 459--474.

\bibitem{60}
L.~Valenta and B.~Rowan, ``{Blindcoin: Blinded, Accountable Mixes for
  Bitcoin},'' in \emph{Int. Conf. Financ. Cryptogr. Data Secur.}\hskip 1em plus
  0.5em minus 0.4em\relax Springer Berlin Heidelberg, 2015, pp. 112--126.

\bibitem{66}
J.~H. Ziegeldorf, F.~Grossmann, M.~Henze, N.~Inden, and K.~Wehrle,
  ``{Coinparty: Secure multi-party mixing of bitcoins},'' in \emph{Proc. 5th
  ACM Conf. Data Appl. Secur. Priv. - CODASPY '15}.\hskip 1em plus 0.5em minus
  0.4em\relax New York, New York, USA: ACM Press, 2015, pp. 75--86.

\bibitem{62}
E.~Heilman, F.~Baldimtsi, and S.~Goldberg, ``{Blindly Signed Contracts:
  Anonymous On-Blockchain and Off-Blockchain Bitcoin Transactions},'' in
  \emph{Int. Conf. Financ. Cryptogr. Data Secur.}\hskip 1em plus 0.5em minus
  0.4em\relax Springer Berlin Heidelberg, 2016, pp. 43--60.

\bibitem{63}
E.~Heilman, L.~AlShenibr, F.~Baldimtsi, A.~Scafuro, and S.~Goldberg,
  ``{TumbleBit: An Untrusted Bitcoin-Compatible Anonymous Payment Hub},'' in
  \emph{Proc. 2017 Netw. Distrib. Syst. Secur. Symp.}\hskip 1em plus 0.5em
  minus 0.4em\relax Reston, VA: Internet Society, 2017.

\bibitem{37}
K.~Kanemura, K.~Toyoda, and T.~Ohtsuki, ``{Design of privacy-preserving mobile
  Bitcoin client based on $\gamma$-deniability enabled bloom filter},'' in
  \emph{2017 IEEE 28th Annu. Int. Symp. Pers. Indoor, Mob. Radio Commun.}\hskip
  1em plus 0.5em minus 0.4em\relax IEEE, oct 2017, pp. 1--6.

\bibitem{42}
H.~Wang, D.~He, and Y.~Ji, ``{Designated-verifier proof of assets for bitcoin
  exchange using elliptic curve cryptography},'' \emph{Futur. Gener. Comput.
  Syst.}, jul 2017.

\bibitem{34}
M.~C.~K. Khalilov and A.~Levi, ``{A Survey on Anonymity and Privacy in
  Bitcoin-like Digital Cash Systems},'' \emph{IEEE Commun. Surv. Tutorials},
  pp. 1--1, 2018.

\bibitem{44}
B.~Qin, J.~Huang, Q.~Wang, X.~Luo, B.~Liang, and W.~Shi, ``{Cecoin: A
  decentralized PKI mitigating MitM attacks},'' \emph{Futur. Gener. Comput.
  Syst.}, oct 2017.

\bibitem{12}
J.-H. Lee, ``{BIDaaS: Blockchain Based ID As a Service},'' \emph{IEEE Access},
  vol.~6, pp. 2274--2278, 2018.

\bibitem{29}
Q.~Lin, H.~Yan, Z.~Huang, W.~Chen, J.~Shen, and Y.~Tang, ``{An ID-based
  linearly homomorphic signature scheme and its application in blockchain},''
  \emph{IEEE Access}, pp. 1--1, 2018.

\bibitem{53}
C.~Yang, X.~Chen, and Y.~Xiang, ``{Blockchain-based publicly verifiable data
  deletion scheme for cloud storage},'' \emph{J. Netw. Comput. Appl.}, vol.
  103, pp. 185--193, feb 2018.

\bibitem{54}
\BIBentryALTinterwordspacing
Y.~Hu, A.~Manzoor, P.~Ekparinya, M.~Liyanage, K.~Thilakarathna, G.~Jourjon,
  A.~Seneviratne, and M.~E. Ylianttila, ``{A Delay-Tolerant Payment Scheme
  Based on the Ethereum Blockchain},'' jan 2018. [Online]. Available:
  \url{http://arxiv.org/abs/1801.10295}
\BIBentrySTDinterwordspacing

\bibitem{58}
C.~Lin, D.~He, X.~Huang, M.~K. Khan, and K.-K.~R. Choo, ``{A New Transitively
  Closed Undirected Graph Authentication Scheme for Blockchain-based Identity
  Management Systems},'' \emph{IEEE Access}, pp. 1--1, 2018.

\bibitem{boireau2018securing}
O.~Boireau, ``Securing the blockchain against hackers,'' \emph{Network
  Security}, vol. 2018, no.~1, pp. 8--11, 2018.

\bibitem{orwl}
``This ultra-secure pc self destructs if someone messes with it,''
  \url{https://www.wired.com/2017/06/orwl-secure-desktop-computer/}, accessed:
  2018-06-01.

\bibitem{lind2017teechain}
J.~Lind, I.~Eyal, F.~Kelbert, O.~Naor, P.~Pietzuch, and E.~G. Sirer,
  ``Teechain: Scalable blockchain payments using trusted execution
  environments,'' \emph{arXiv preprint arXiv:1707.05454}, 2017.

\bibitem{bentov2017tesseract}
I.~Bentov, Y.~Ji, F.~Zhang, Y.~Li, X.~Zhao, L.~Breidenbach, P.~Daian, and
  A.~Juels, ``Tesseract: Real-time cryptocurrency exchange using trusted
  hardware,'' 2017.

\bibitem{PoS2015}
S.~Dziembowski, S.~Faust, V.~Kolmogorov, and K.~Pietrzak, ``Proofs of space,''
  in \emph{Proc. 35th Annual Cryptology Conference on Advances in Cryptology},
  Aug. 2015, pp. 585--605.

\bibitem{DPos}
\BIBentryALTinterwordspacing
``{DPOS} description on bitshares,'' accessed on 15 June, 2018. [Online].
  Available: \url{http://docs.bitshares.org/ bitshares/dpos.html}
\BIBentrySTDinterwordspacing

\bibitem{ProofofStake}
\BIBentryALTinterwordspacing
``Telehash,'' accessed on 15 June, 2018. [Online]. Available:
  \url{http://telehash.org}
\BIBentrySTDinterwordspacing

\bibitem{FrancaMiniBlock2015}
\BIBentryALTinterwordspacing
B.~F. Fran\c{c}a, ``Homomorphic mini-blockchain scheme,'' pp. 1--17, Apr. 2015,
  accessed on 15 June, 2018. [Online]. Available:
  \url{http://cryptonite.info/files/HMBC.pdf}
\BIBentrySTDinterwordspacing

\bibitem{Bruce2014}
\BIBentryALTinterwordspacing
J.~D. Bruce, ``The mini-blockchain scheme,'' 2014, accessed on 15 June, 2018.
  [Online]. Available: \url{http://www.cryptonite.
  info/files/mbc-scheme-rev2.pdf}
\BIBentrySTDinterwordspacing

\bibitem{ayres2016cyberterrorism}
N.~Ayres and L.~A. Maglaras, ``Cyberterrorism targeting the general public
  through social media,'' \emph{Security and Communication Networks}, vol.~9,
  no.~15, pp. 2864--2875, 2016.

\bibitem{chen2018towards}
Y.~Chen, Q.~Li, and H.~Wang, ``Towards trusted social networks with blockchain
  technology,'' \emph{arXiv preprint arXiv:1801.02796}, 2018.

\bibitem{kong2018achieving}
Q.~Kong, R.~Lu, H.~Zhu, and M.~Ma, ``Achieving secure and privacy-preserving
  incentive in vehicular cloud advertisement dissemination,'' \emph{IEEE
  Access}, vol.~6, pp. 25\,040--25\,050, 2018.

\bibitem{hua2018cinema}
J.~Hua, H.~Zhu, F.~Wang, X.~Liu, R.~Lu, H.~Li, and Y.~Zhang, ``Cinema:
  Efficient and privacy-preserving online medical primary diagnosis with
  skyline query,'' \emph{IEEE Internet of Things Journal}, pp. 1--1, 2018.

\end{thebibliography}

\end{document}